\documentclass[twocolumn,aps,prl,superscriptaddress]{revtex4-2}  
\usepackage{graphicx}  
\usepackage{bm}        
\usepackage{amssymb}   
\usepackage{xcolor}
\usepackage{amsmath}
\usepackage{physics}
\usepackage{tabularx}
\usepackage{bm}
\usepackage{placeins}
\usepackage[textheight=672pt,textwidth=510pt]{geometry}

\usepackage{layout}
\RequirePackage{totcount,xpatch}
\definecolor{linkColor}{RGB}{0,80,150}
\usepackage{hyperref}
\hypersetup{
    colorlinks=true,
    allcolors=linkColor,
    pdfborder={0 0 0},
    pdfencoding = auto
}

\usepackage[normalem]{ulem}

\newcommand{\Qb}{\mathbf{Q}}
\newtotcounter{movie}
\renewcommand{\themovie}{S\arabic{movie}}
\newcommand{\movie}[1]{%
  \refstepcounter{movie}%
  \noindent\textbf{Movie \themovie.\space#1}\par\bigskip
}

\AtBeginEnvironment{tabular}{
\sffamily\fontsize{7.5}{10}\selectfont
}

\begin{document}

\title{Asymmetric fluctuations and self-folding of active interfaces}

\author{Liang Zhao$^{1*}$, Paarth Gulati$^{1*}$, Fernando Caballero$^{1,2}$, Itamar Kolvin$^{1,3}$, Raymond Adkins$^{1,4}$,\\ M. Cristina Marchetti$^{1,5}$, Zvonimir Dogic$^{1,5}$\\ \normalsize $^{1}$\textit{Department of Physics, University of California, Santa Barbara, Santa Barbara, California 93106, USA}\\ \normalsize $^{2}$\textit{Department of Physics, Brandeis University, Waltham, MA 02453, USA} \\ \normalsize 
$^{3}$\textit{School of Physics, Georgia Institute of Technology, Atlanta, GA 30332, USA} \\ \normalsize 
$^{4}$\textit{Department of Molecular Biophysics and Biochemistry, Yale University, New Haven, CT 06520, USA}\\ \normalsize 
$^{5}$\textit{Biomolecular Science and Engineering Program, University of California, Santa Barbara, CA 93106, USA}\\  
\small{Corresponding authors: mcmarche@ucsb.edu (M.C.M.), zdogic@ucsb.edu (Z.D.)}\\ \small{$^{*}$These two authors contributed equally} \\ }

\begin{abstract}
We study the structure and dynamics of the interface separating a passive fluid from a microtubule-based active fluid. Turbulent-like active flows power giant interfacial fluctuations, which exhibit pronounced asymmetry between regions of positive and negative curvature. Experiments, numerical simulations, and theoretical arguments reveal how the interface breaks up the spatial symmetry of the fundamental bend instability to generate local vortical flows that lead to asymmetric interface fluctuations. The magnitude of interface deformations increases with activity: In the high activity limit, the interface self-folds invaginating passive droplets and generating a foam-like phase, where active fluid is perforated with passive droplets. These results demonstrate how active stresses control the structure, dynamics, and break-up of soft, deformable, and reconfigurable liquid-liquid interfaces.
\end{abstract}

\maketitle


Bulk active fluids composed of extensile rods are inherently unstable~\cite{Simha2002}. An initially uniformly aligned state rapidly evolves into steady-state turbulent-like dynamics through the amplification of infinitesimal bend distortions by the active stresses ~\cite{Sanchez,giomi2014defect,zhou2014living,thampi2014instabilities,giomi2015geometry,gao2015multiscale,guillamat2016control,Martinez-Prat2019,alert2020universal,kumar2018tunable,PoojaPRL,Sokolov_PRX,head2024spontaneous}. Active fluids exhibit a complex interplay with rigid boundaries, geometrical confinements, and deformable interfaces or membranes. On the one hand, rigid and immutable boundaries dramatically change the structure and dynamics of a bulk-active fluid. For example, geometrical confinement transforms bulk chaotic dynamics into coherent long-ranged flows capable of efficient mass transport \cite{Wioland_PRL,Nishiguchi_NatureC,Kun-ta_science, Opathalage_PNAS,theillard2017geometric,shendruk2017dancing,varghese2020confinement,chandragiri2020flow,gulati2022boundaries}. On the other hand, the active fluid can exert stresses on soft and deformable interfaces and membranes, regulating their structure and dynamics and generating conformations and morphologies that are not accessible in equilibrium. For example, in biological cells, a deformable lipid membrane interacts with the stress-generating actin cortex to enable essential processes, such as blebbing, endo and exo-cytosis, cell division and wound-healing~\cite{tinevez2009role,Sedzinski2011,Salbreux_PRL,reymann2016cortical}. It is, however, challenging to elucidate universal physical principles in complex biological cells. Progress can instead be made by studying simplified systems~\cite{joanny2012drop,tjhung2012spontaneous,blow2014biphasic,giomi2014spontaneous,paoluzzi2016shape,miles2019active,soni2019stability,ray,young2021many,kempf2019active,alert2022fingering,coelho2020propagation,xu2023geometrical,ruske2021morphology}. 

So far, experiments have mostly explored active fluids confined within finite-sized lipid vesicles or droplets~\cite{Keber_science,Takatori2020,Vutukuri2020,Velez-Ceron_PNAS,kokot2022spontaneous,Tayar_NatureP,sciortino2023active}. We study microtubule-based active fluids that exhibit long-ranged turbulent-like flows and can be prepared in milliliter quantities~\cite{Sanchez}. Using a recently developed system~\cite{ray}, we explore how such bulk fluids interact with macroscale deformable and reconfigurable interfaces that are created by liquid-liquid phase separation. The active fluid partitions into one of the two phases, generating active stresses that power dramatic interfacial fluctuations. Active interfaces exhibit a pronounced up-down asymmetry in the local curvature, which is explained by how the interface breaks the spatial symmetry of the ubiquitous bend instability. Increasing the activity beyond a critical value leads to giant fluctuations where the interface folds on itself, enveloping passive droplets and generating foam-like perforated phases.

\section*{Experimental realization of a bulk active interface}

\FloatBarrier
\begin{figure*}[t]
\centering
\includegraphics[width=0.7\textwidth]{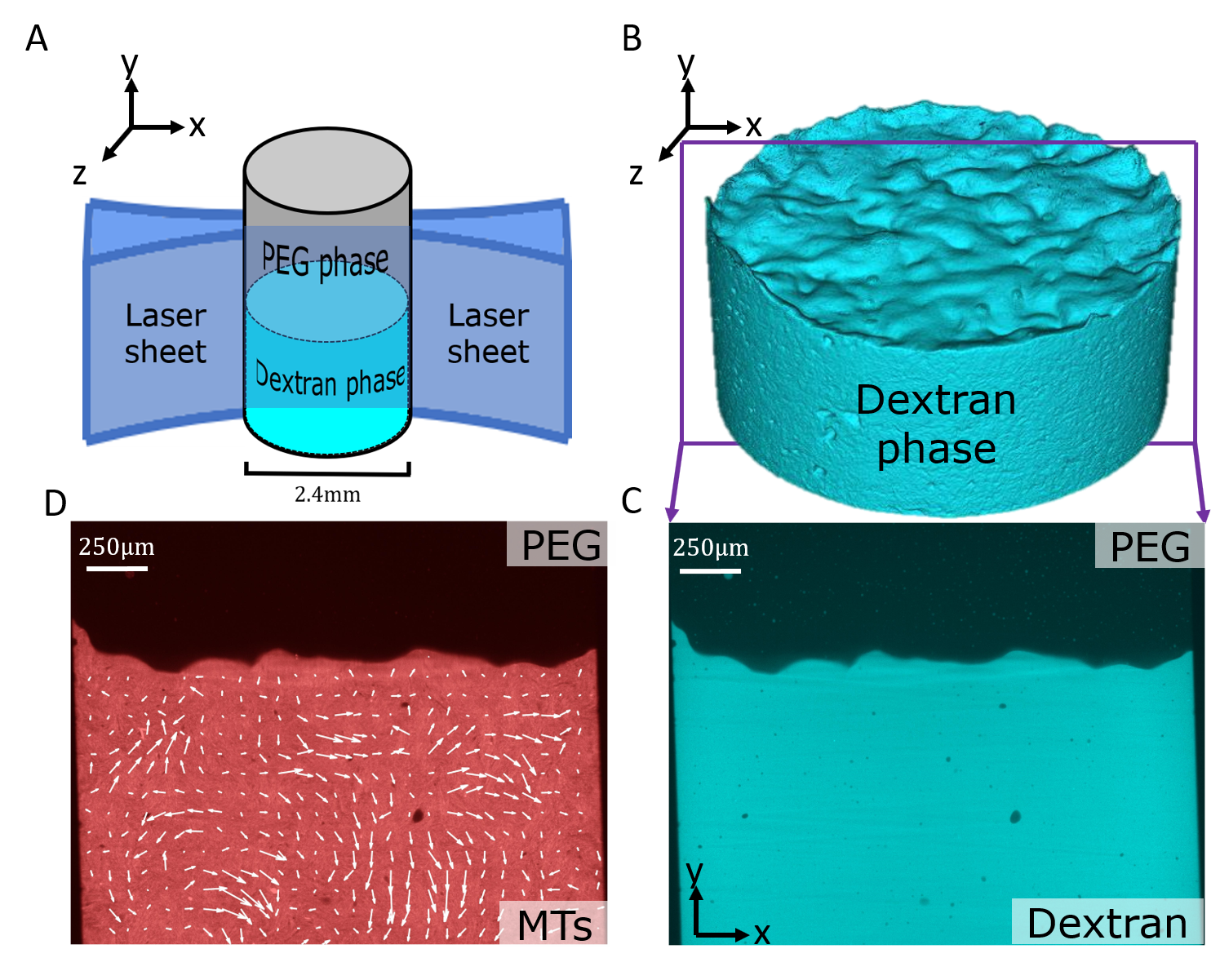}
\caption{{Bulk phase separation between an active and a passive phase.} (A) Experimental setup used for imaging the active interface. The sample was loaded into a transparent cylindrical tube, where it separated into a passive top PEG-rich layer and a bottom dextran-rich layer (cyan) that contained the MT-based active fluid. The sample was illuminated with a laser sheet. The $x-y$ plane was imaged by an objective aligned along the $z$ direction. The tube's inner diameter was 2.4~mm. (B) Fluorescent dextran phase, reconstructed from 2D $x-y$ images at 120 different $z$ positions. (C) A vertical $x-y$ slice in the middle of the cylinder. Dextran is indicated in cyan. (D) The same slice was imaged in the MT (red) channel. MTs strongly partition with the dextran phase. White arrows are the 2D velocity field. The sample contained 129 nM KSA, 2.1\% PEG and 2.1\% dextran.}
\label{setup}
\end{figure*}
The passive phase-separating system was a mixture of polyethylene glycol (PEG, 100~kDa) and dextran (500~kDa) dissolved in an aqueous buffer. At high polymer concentration, this mixture separated into a PEG-rich and a dextran-rich phase \cite{Yonggang_Langmuir}. We merged the PEG/dextran mixture with an active fluid, which contained MTs and force-generating kinesin-streptavidin clusters (KSA)~\cite{ray}. The active fluid components partitioned into the dextran phase where MTs bundled due to the depletion attraction (Fig.~\ref{setup}D)~\cite{Hilitski}. The KSA clusters crosslinked multiple MTs within a bundle. The ATP-fueled stepping of kinesin motors drove interfilament sliding, which generated extensile active stresses that powered turbulent-like dynamics throughout the dextran phase. An ATP-regeneration system sustained the active dynamics for several hours. The PEG/dextran concentrations were adjusted to match the interfacial tension with the active stresses~\cite{Yonggang_Langmuir}. Two systems were studied: a ``high'' interfacial tension mixture of 2.1\% PEG and 2.1\% dextran, and a ``low'' interfacial tension mixture of 2.0\% PEG and 2.0\% dextran (Fig.~S11).

Active interfaces and associated phase separation were previously studied in a quasi-2D confinement \cite{ray}. In this regime, the strong frictional interaction with the confining walls suppressed the interface mobility and limited the magnitude and range of active flows. Additionally, the interface exhibited activity-induced wetting, which spread the active phase along the confining walls and changed the interface structure. To minimize boundary effects and frictional interactions, we created a 2D active interface that separated a bulk 3D active fluid and a 3D passive fluid. The active mixture was placed in a cylindrical chamber, where it spontaneously phase-separated into PEG-rich and dextran-rich components (Fig.~\ref{setup}A). Centrifugation resulted in bulk phase separation layering the PEG-rich phase on top of the denser active dextran-rich phase. A fraction of the dextran and the MTs were fluorescently tagged. Samples were imaged with a light sheet microscope (Zeiss Z.1 Lightsheet). Volumes were reconstructed from $x-y$ cross-sections obtained at successive $z$ positions with a step size of 20 $\mu$m (Fig.~\ref {setup} B). Scanning a 200 $\mu$m thick section in two channels required 3.5~s and scanning a 600~$\mu$m section in one channel required 4.3~s.

\begin{figure*}[t]
    \centering
    \includegraphics[width=\textwidth]{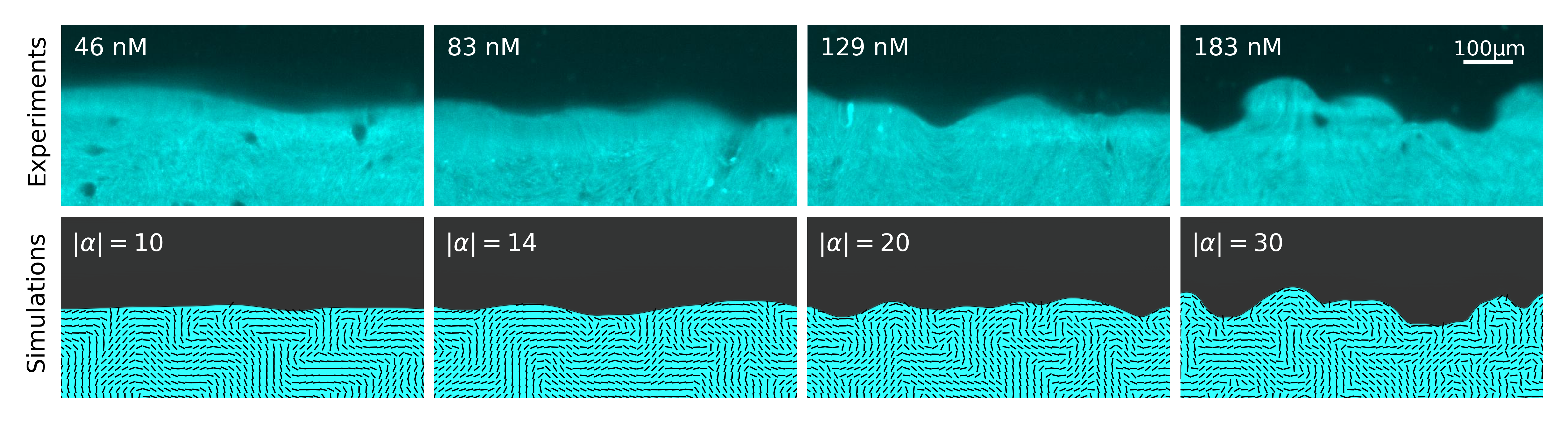}
    \caption{Asymmetric interfacial fluctuations in experiments and simulations. \textbf{Top row (Experiments):} Two-dimensional cross-sections of the interface at increasing KSA concentrations. The amplitude of fluctuations increases with increasing KSA concentration. All samples contained 2.1\% PEG and 2.1\% dextran. The images were taken at $t=2$~h. \textbf{Bottom row (Simulations):} Both the amplitude and asymmetry of the interfacial fluctuations increase with activity at a fixed surface tension, $\gamma=500$. The passive and active phases are shown in black and cyan, respectively. The nematic director inside the active phase is plotted every 30 lattice sites. Only a portion of both experimental and simulation systems close to the bulk interface is shown here. }
    \label{raw_activity}
\end{figure*}

Immediately after centrifuging, the interface was flat. Spontaneous flows developed in the active phase generated interfacial deformations on the scale of 10's of microns (Fig.~\ref{setup} B-D)~\cite{ray}. The amplitude of interfacial fluctuations, which is a proxy to the active forces operating at the interface, increased with time (Fig. S10, Movie S1). The magnitude of interfacial fluctuations was correlated with the KSA concentration in the range of 45 nM to 180 nM (Fig.~\ref{raw_activity}, Movie S2). 

\section*{Continuum model of active phase separation}
\label{sec:ContinuumModel}

To gain a qualitative understanding of the experimental findings, we explored a numerical model of active interfaces~\cite{blow2014biphasic,ray,Tayar_NatureP,Fernando}. The model was implemented in two dimensions, with a one-dimensional interface separating the active and passive phases without frictional dissipation. The hydrodynamic equations were numerically solved for a mixture of an active liquid crystal and a passive fluid, given by
\begin{equation}
\begin{aligned}
    D_t \phi &= M \nabla^2 \mu,\\
    D_t \Qb &= \lambda \mathbf{A} - \boldsymbol{\omega}\cdot \Qb + \Qb \cdot \boldsymbol{\omega} + \frac{1}{\Gamma} \mathbf{H},\\
    0 &= \eta\nabla^2 \mathbf{v} - \nabla P + \nabla \cdot \boldsymbol{\sigma} + \mathbf{f}^g,
    \label{eq:continuum_model}
\end{aligned}
\end{equation}
where $\phi$ is the concentration of active fluid, the nematic tensor $\Qb$ describes the local orientational order in the active liquid crystal, and $\mathbf{v}$ is the flow velocity which advects both $\phi$ and $\Qb$ through the material derivative $D_t = \partial_t + \mathbf{v} \cdot \nabla$. Phase separation between the active ($\phi=1)$ and passive $(\phi=0)$ phases is driven by the chemical potential $\mu = \delta F_\phi/\delta \phi$, where $F_\phi$ is the Cahn-Hilliard free energy, $F_\phi = ({4\gamma}/{\xi})\int d\mathbf{r} \left( \phi^2 (\phi-1)^2 + (\xi^2/2) (\nabla\phi)^2\right)$. Here, $\gamma$ is the surface tension of the phase-separated mixture and $\xi$ is the interface width.

The nematic order parameter $\Qb$ is rotated by flow gradients quantified by the strain rate $A_{ij} = (\partial_i v_j + \partial_j v_i)/2$ and vorticity $\omega_{ij} = (\partial_i v_j - \partial_j v_i)/2$ and relaxes through $\mathbf{H} =- \delta F_Q/\delta \Qb$, where $F_Q$ is the Landau-de Gennes free energy for the mixture, $F_Q  =  \int d\mathbf{r} (r\phi/2) \text{Tr}{\Qb^2} + (u/4) (\text{Tr}{\Qb^2})^2 + (K/2) (\nabla \Qb)^2$, with $\Gamma$ a rotational viscosity. Here $K$ is the nematic elastic modulus and $u>0$. To model the experimental system, we choose $r>0$, which ensures that in the absence of activity, the liquid crystal is in the isotropic phase.

The flow is governed by Stokes' equation and assumed incompressible, $\nabla \cdot \mathbf{v} =0$. It is controlled by the interplay of viscous dissipation with viscosity $\eta$, pressure gradients, and passive and active stresses, $\bm\sigma=\bm\sigma^p+\bm\sigma^a$. The active stress $\bm\sigma^a = \alpha\phi \Qb$ is controlled by the fraction $\phi$ of active fluid. The extensile nature of the liquid crystal is described by the activity $\alpha<0$. Finally, $\mathbf{f}_g =  - \hat{y}(\rho_0 + \phi \Delta \rho) g $ is the force due to gravity, with $\Delta \rho$ the difference in the densities of the two fluid phases. More details about the model are given in the SI.

Active flows generate local nematic order through flow alignment~\cite{Kun-ta_science,srivastava2016negative,santhosh2020activity}. Above a critical activity $\alpha_0=2\eta r/\lambda \Gamma$, bulk isotropic active liquid crystals are linearly unstable and develop spatiotemporal chaotic flows similar to those observed in the nematic phase~\cite{giomi2014spontaneous}. We consider activities with magnitudes larger than $\alpha_0$. We vary the surface tension $\gamma$, keeping the interface width fixed. We measure lengths in units of the interface width $\xi$, times in units of the nematic reorientation time $\Gamma/r$, and stresses in units of liquid crystal energy density $r.$ In these units, the other parameters in the model are fixed as follows: $u =10.0, M =0.67, K=106.67, \eta=1.0, \lambda=1.0$, and $\Delta \rho g = 0.12,$  which correspond to $\alpha_0=2.0$. We solve eqs.~\ref{eq:continuum_model} numerically using self-developed pseudo-spectral solvers. All simulation results are shown for a regular grid with $1024\times 1024$ lattice sites, and a total of $10^6-10^7$ time steps. Details about the numerical simulations are in Materials and Methods and the SI. 

\section*{Asymmetry of fluctuating active interfaces}
\begin{figure}[t]
    \centering    
    \includegraphics[width=0.49\textwidth]{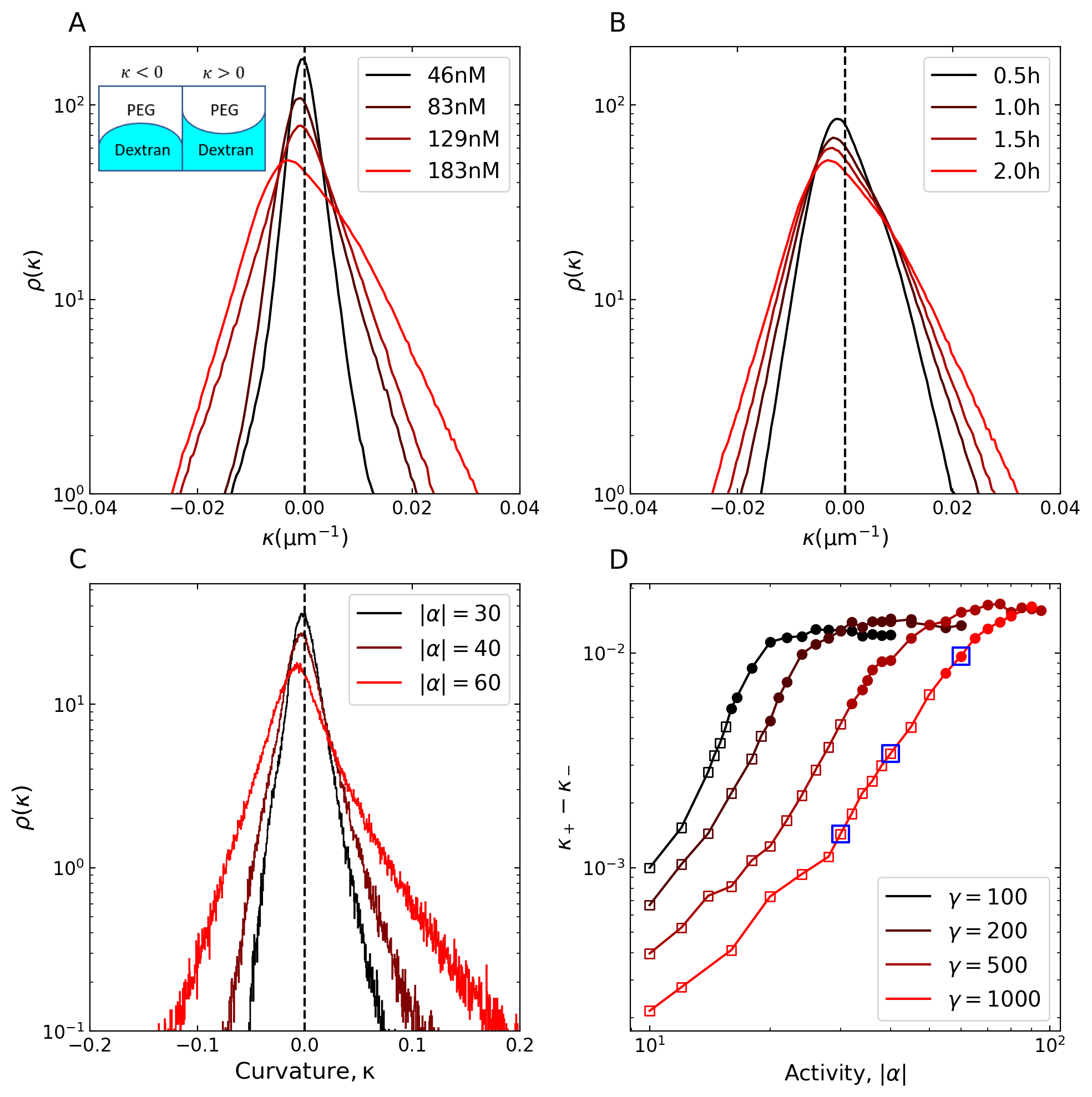}
    \caption{Asymmetry in the probability distribution of local interface curvatures $\rho(\kappa)$. \textbf{Top row (Experiments):} (A) Curvature probability distribution $\rho(\kappa)$ for four KSA concentrations at time $t=2$~h. Inset: defining $\kappa$ sign. (B) Time evolution of $\rho(\kappa)$ for 183~nM KSA sample. Each curve is averaged over 30 minutes. All samples contain 2.1\% PEG and 2.1\% dextran. \textbf{Bottom row (Simulations):} (C) Curvature probability distribution at three values of activity  $\alpha$ and fixed surface tension $\gamma=1000$. (D) The curvature asymmetry $\kappa_+-\kappa_-$ ( Eqs.~\ref{eq:kappa_pm}) increases with increasing activity and with decreasing surface tension $\gamma$. The asymmetry saturates at large activity, where the interface breaks up due to the invagination of droplets of passive fluid into the active phase. Open and closed symbols correspond to states without and with droplet invagination. The blue squares correspond to the curvature distributions in panel C.}
    \label{fig:curvature_exp_sim}
\end{figure}

We first studied the ``high''-surface-tension regime, where interfaces retained their integrity throughout most of the experiment. Careful inspection revealed a peculiar feature of such interfaces. Typical interfacial configurations, especially at higher activity, were characterized by steep and deep valleys wherein the passive phase protruded into the active phase (Fig.~\ref{raw_activity}A). In contrast, the peaks of the active phase into the passive phase were shallow and smooth. Motivated by these observations, we calculated the local mean interface curvatures $\kappa$ and plotted their distribution over the whole interface (Fig.~\ref{fig:curvature_exp_sim}A) (Materials and Methods). Positive curvatures indicated valleys where the passive phase protrudes into the active dextran phase (Fig.~\ref{fig:curvature_exp_sim}A inset). The width of the distribution increased with increasing KSA concentrations, which controls sample activity (Fig.~\ref{fig:curvature_exp_sim}A). Additionally, the probability density of local curvature exhibited a pronounced asymmetry. For large amplitudes, the probability of observing positive curvatures was larger than the negative ones ($\rho(|\kappa|)>\rho(-|\kappa|)$). Since the distribution was measured along a plane, its average must be zero. Therefore, the asymmetric curvature distributions did not peak at $\kappa=0$, but at a small negative value of $\kappa$. This quantification confirmed that the protrusions of the passive phase into the active phase were sharper and more pronounced than the reverse.

The interface reconfigures on a time scale of tens of seconds. Thus, over tens of minutes, it undergoes a sufficient number of independent conformations to allow for accurate measurement of the curvature distribution. Plotting the time evolution of such distributions over the sample lifetime reveals that the interface exhibited an intriguing aging effect (Fig.~\ref{fig:curvature_exp_sim}B, S10, Movie S1). The magnitude of the fluctuations and the width of the curvature distributions increased over time. This trend was similar for interfaces measured at a fixed time but with an increasing KSA concentration (Fig. S8, Movie S2). It was also reported in the previous study of quasi-2D interfaces~\cite{ray}. These observations suggest that in MT-based active fluids the active stress increases with time. The origin of this behavior is unknown.  

To understand the asymmetry mechanism we examined the behavior of the interface in the continuum model. The interface is defined as the locus of points corresponding to where the polymer concentration $\phi=1/2$. We calculated the local curvature $\kappa(s)$ at the arclength $s$ along the interface after the numerically integrated equations reached a steady state (Fig. S5). The probability distribution of $\kappa(s)$ is asymmetric (Fig. \ref{fig:curvature_exp_sim}C), as found in experiments, with longer tails for $\kappa >0$ than for $\kappa<0$, corresponding to smooth peaks and deep valleys. The asymmetry is quantified by the difference between mean positive and negative curvatures, defined as
\begin{equation}
    \kappa_+ = \dfrac{\int_0^\infty d\kappa\; \kappa~ \rho(\kappa)}{\int_0^\infty d\kappa\; \rho(\kappa)}; \;\;\; \kappa_- = \left| \dfrac{\int_{-\infty}^0d\kappa\; \kappa~ \rho(\kappa)}{\int_{-\infty}^0 d\kappa\; \rho(\kappa)} \right|\;.
    \label{eq:kappa_pm}
\end{equation}
For symmetric fluctuations, $\kappa_+ - \kappa_- = 0$. We find that this difference increases with activity, but saturates at high activity (Fig.~\ref{fig:curvature_exp_sim}D).

\begin{figure}[t]
    \centering
    \includegraphics[width=0.49\textwidth]{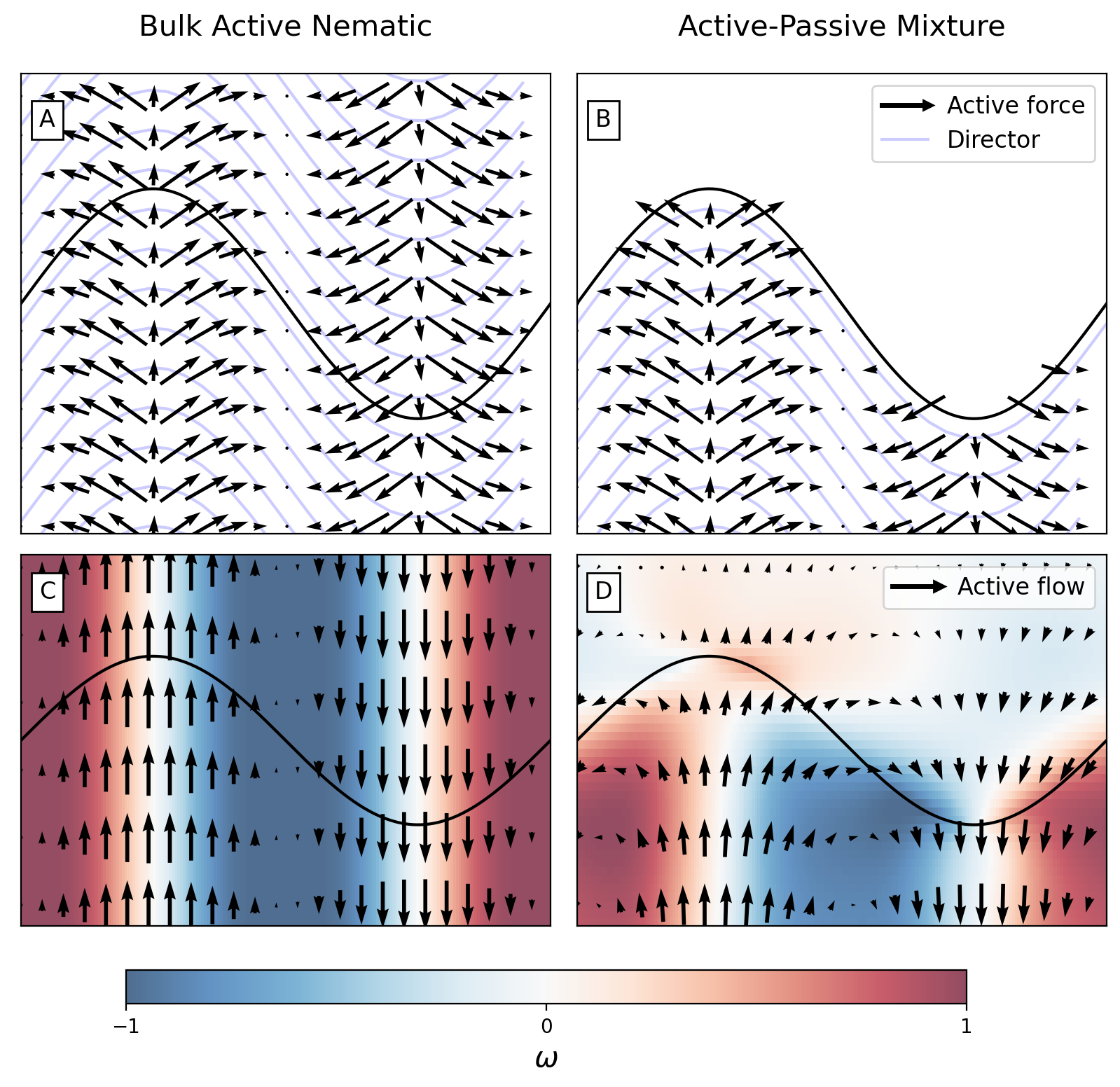}
    \caption{Active forces and flows in a bulk active fluid and in a phase-separated mixture. (A) Analytically calculated active forces from a sinusoidal deformation of the director field in a bulk active fluid (Eqs.~S15). Black arrows are active forces and light blue lines are the director field. The black solid line is a guide to the eye as there is no interface here. (B) Active forces from director deformations in a phase-separated active/passive mixture (Eqs.~S14). The black solid line denotes the interface between passive (top) and active (bottom) fluids. (C) Active flows are obtained from the numerical solution of the Stokes equation for a bulk active fluid with the nematic configuration in the panel above. The color map is vorticity $\omega$. (D) In the phase-separated mixture, active forces are absent from the passive phase, leading to asymmetric tangential flows that deepen the valleys and broaden the peaks.}   
    \label{fig:interface_flow_sketch}
\end{figure}
We hypothesize that the asymmetry arises from the interplay between the interface-induced structures of the active fluid and the flows they generate. In aligned nematics, extensile activity creates effective interaction that aligns the director field with interfaces or walls~\cite{blow2014biphasic,coelho2023active}. We find that the same phenomenon, known as active anchoring, occurs in the isotropic active phase of extensile active liquid crystals (Fig. S3). The thickness of the aligned layer is controlled by the active length scale $\ell_{\alpha} \sim \sqrt{K/|\alpha|}$ (Fig. S4). Given the presence of a surface-induced aligned layer, we compare active forces and associated flows in a phase-separated active/passive mixture to those in a bulk active fluid.  We analytically calculate the active forces induced by director deformations (Fig.~\ref{fig:interface_flow_sketch}A, B) and the associated active flows obtained from the Stokes equation (Fig.~\ref{fig:interface_flow_sketch}C, D, SI Sec. 2A). In a bulk active fluid, active forces yield the familiar bend instability which symmetrically enhances any director bend deformation ~\cite{Simha2002}. In the active/passive mixture, the force-generating active liquid crystal is located on only one side of the interface; active forces are absent in the passive phase. This creates an asymmetry in the flows that deepens and narrows the valleys into the active fluids while stretching and broadening out the shallow peaks (Fig.~\ref{fig:interface_flow_sketch}D). The asymmetry, due to the preferred alignment of the director with the interface, is exaggerated in this idealized limit with perfect anchoring. Similar vortical flows, however, arise due to active anchoring and give rise to curvature asymmetry.

\begin{figure}[t]
    \centering \includegraphics[width=0.49\textwidth]{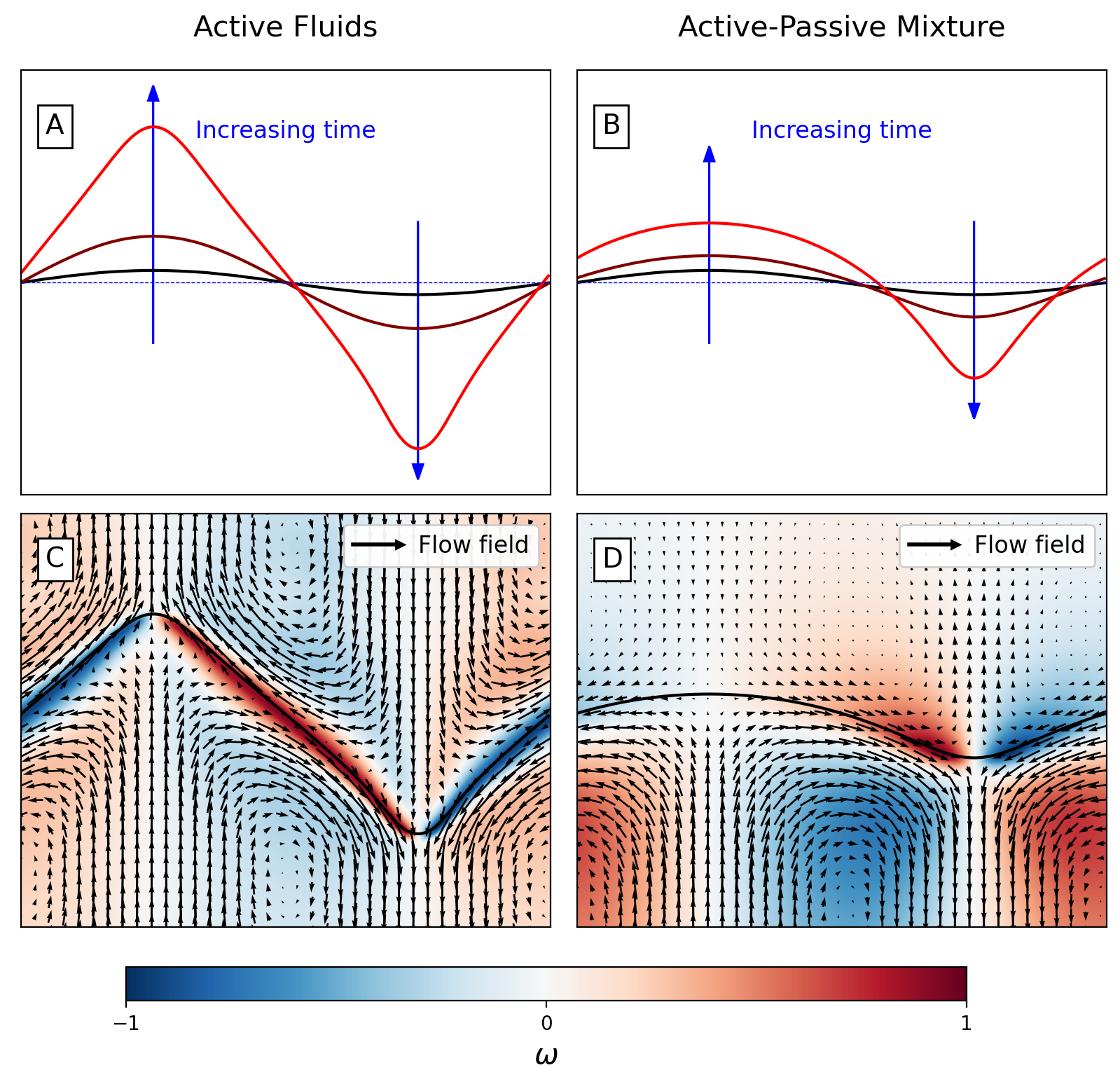}
    \caption{Time evolution of an interface perturbation. The nematic is anchored to the interface, and flows are driven by active and interfacial stresses (SI Section 3B). (A) Time evolution (black to red) of a symmetric interface perturbation in the case of two identical active fluids above and below the interface. The time-evolving interface preserves the initial symmetry (B) Time evolution (black to red) of a symmetric interface perturbation in the case of an active fluid below the interface and a passive isotropic fluid above. The initial symmetric perturbation of the interface builds curvature asymmetry i.e. it develops a deep valley and a broad peak. (C) Flow field near the interface in the case of identical active fluids for the interface configuration shown in red above. Arrows depict the flow velocity and color indicates normalized vorticity $\omega$. (D) Flows near the interface separate the active (below the interface) and passive fluid (above the interface) for the interface configuration shown in red above. The interface experiences asymmetric vortical flows, resulting in curvature asymmetry.}
\label{fig:interface_minimal_sim}
\end{figure}

This dynamical asymmetry becomes more evident by examining the evolution of a sinusoidal perturbation of the interface between a passive fluid and an active nematic with explicit director anchoring (Fig.~\ref{fig:interface_minimal_sim}). In this minimal model, the flow is driven by active forces and relaxes through the passive molecular field $\mathbf{H}$, but does not directly rotate the nematic director (SI Sec.~3B). As the system evolves, the nematic texture reorients to remain parallel to the interface. In a mixture of two identical active nematics, the interface remains symmetric while evolving in time (Fig.~\ref{fig:interface_minimal_sim}A, C). In contrast, in the active/passive mixture (Fig.~\ref{fig:interface_minimal_sim}B, D), the flow deforms the interface asymmetrically, with sharp dips into the active phase.  We note that the asymmetry is a nonlinear effect, driven by flows tangent to the interface. It cannot be recovered by linear stability analysis of a flat or uniformly curved interface that only includes normal active forces~\cite{alert2022fingering}.
 
\section*{Interface Self-folding and Droplet Invagination}
\begin{figure}[t]
    \centering
    \includegraphics[width=0.48\textwidth]{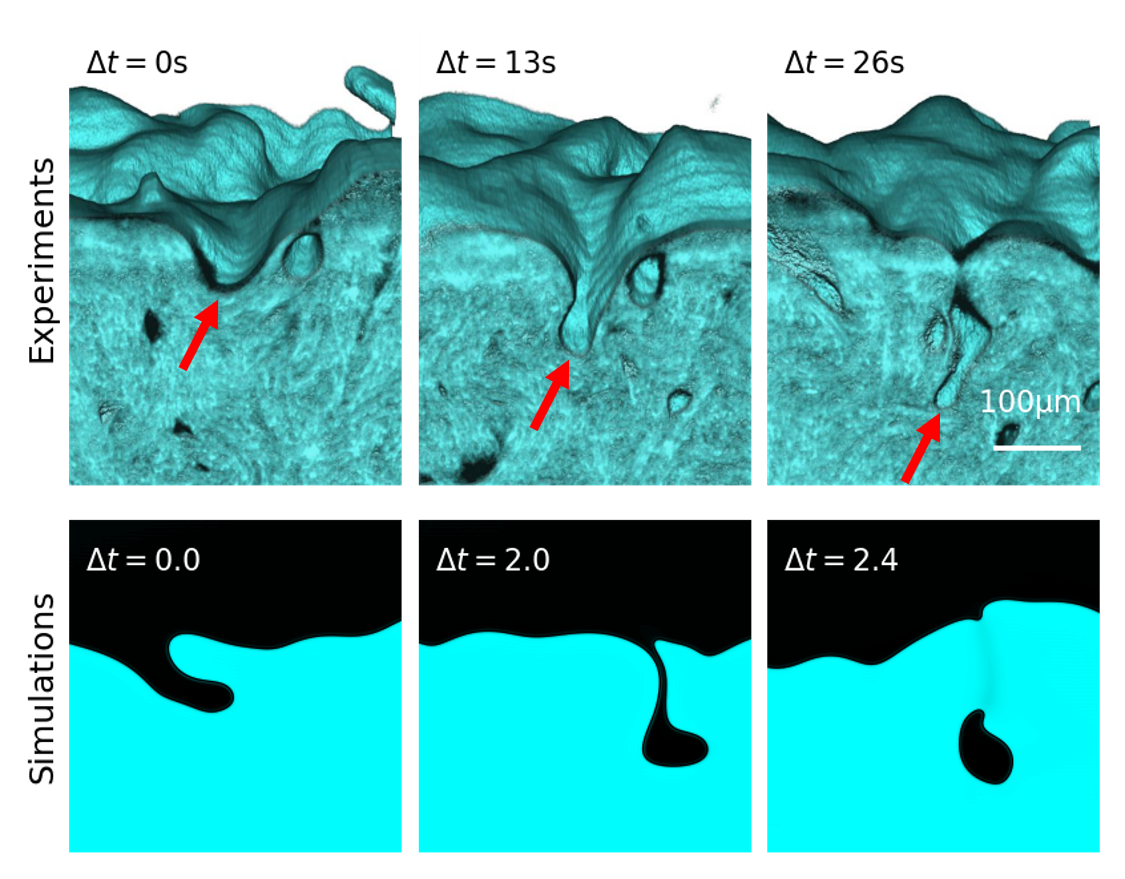}
    \caption{Pathways of passive droplet invagination. \textbf{Top row (Experiments):} Invagination of a passive droplet by the active phase. The valley in the interface grows deeper with time. Eventually, the vertical walls merge to form a passive droplet enveloped by the active fluid (Movie S3). The red arrows mark the position of the invagination. \textbf{Bottom row (Simulations):} Invagination of a passive droplet into the active fluid observed in simulations ($\alpha=-20, \gamma=200$, Movie S8 B)}
    \label{fig:invagination}
\end{figure}

\begin{figure}[t]
\begin{center}
\includegraphics[width=0.47\textwidth]{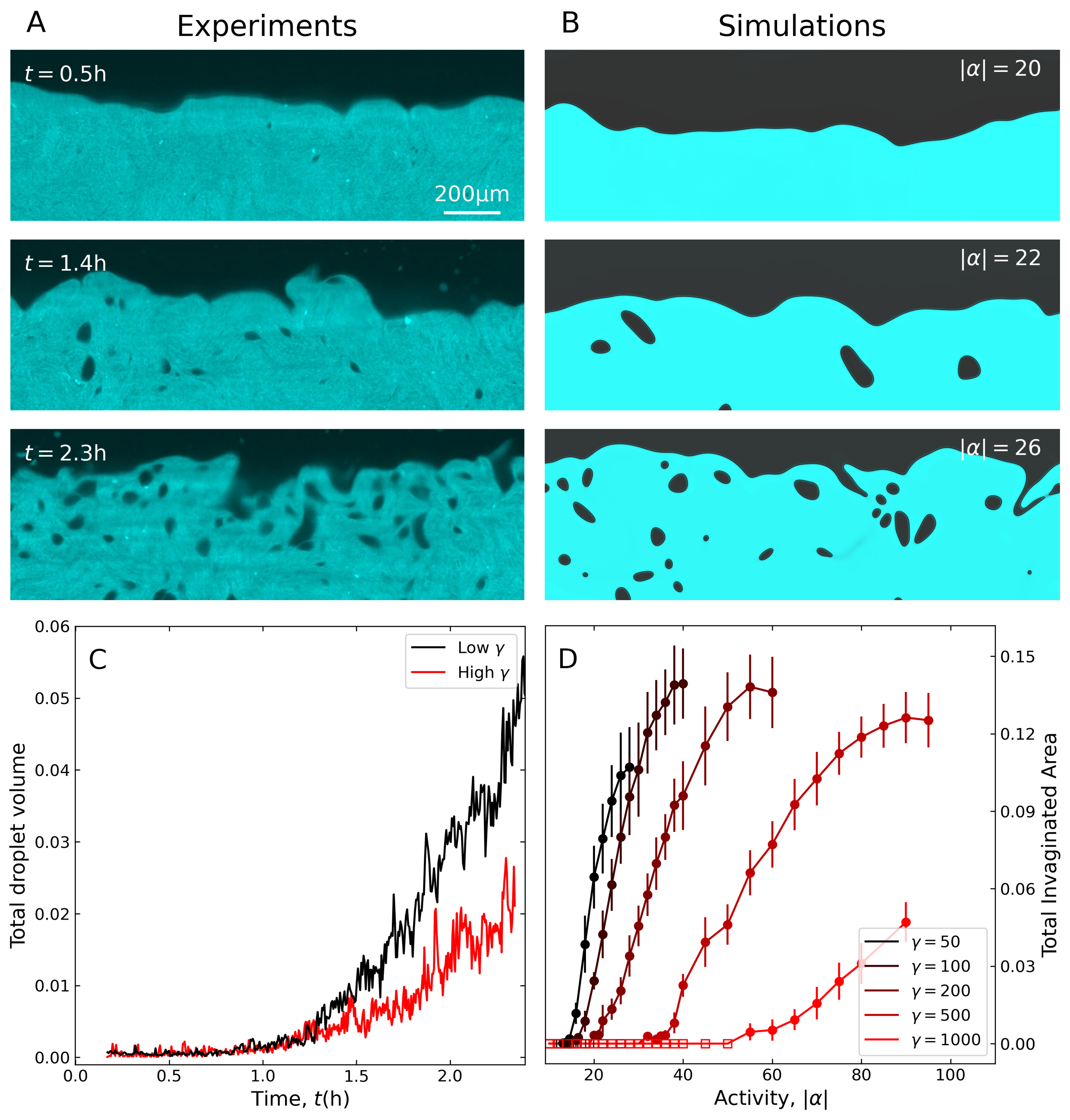}
\end{center}
\caption{Continuous interface breakage and droplet formation. (A). Conformation of a ``low''-tension interface at different times. The sample had 183~nM KSA. (B) Dependence of an interface structure on activity. Surface tension is $\gamma =200.0$. (C) The total volume of enveloped passive droplets as a function of time normalized by the volume of the active phase, for ``low'' and ``high'' surface tension samples. Once the fluctuation increased to a sufficiently large value at $t=1.2$~h, the invagination events took place, and the volume of enveloped droplets increased linearly. (D) Mean total area of the passive droplets in the active phase in steady state (normalized by the total initial amount of active fluid) as a function of activity, for different values of surface tension.}
\label{fig:composite_invaginatedArea}
\end{figure}

Polymer concentration controls the magnitude of the interfacial tension. So far, we studied ``high'' interfacial tension samples where PEG and dextran strongly separate. Decreasing polymer concentration reduced the surface tension, increased the magnitude of the interfacial deformations, and led to a qualitatively new behavior.  When deformations increased above a critical threshold, the interface folded onto itself, invaginating passive droplets into the active phase (Fig.~\ref{fig:invagination}, Movie S3, S4). All such events followed a universal pathway: an initial valley-like interfacial deformation developed into a deep canyon-like structure that grew in depth over time. Eventually, the canyon walls merged, producing a passive droplet entirely enveloped by the active phase. Repeating invagination events created an active phase perforated by passive droplets (Fig.~\ref{fig:composite_invaginatedArea}A, Movie S5-S7). Notably, the break-up of the continuous interface also exhibited a pronounced asymmetry. While we observed frequent invagination of passive droplets within the active phase,  the reverse process where active dextran droplets were ejected into the passive phase was exceedingly rare if ever observed (Movie S6). 

\begin{figure}[h]
\begin{center}
\includegraphics[width=0.49\textwidth]{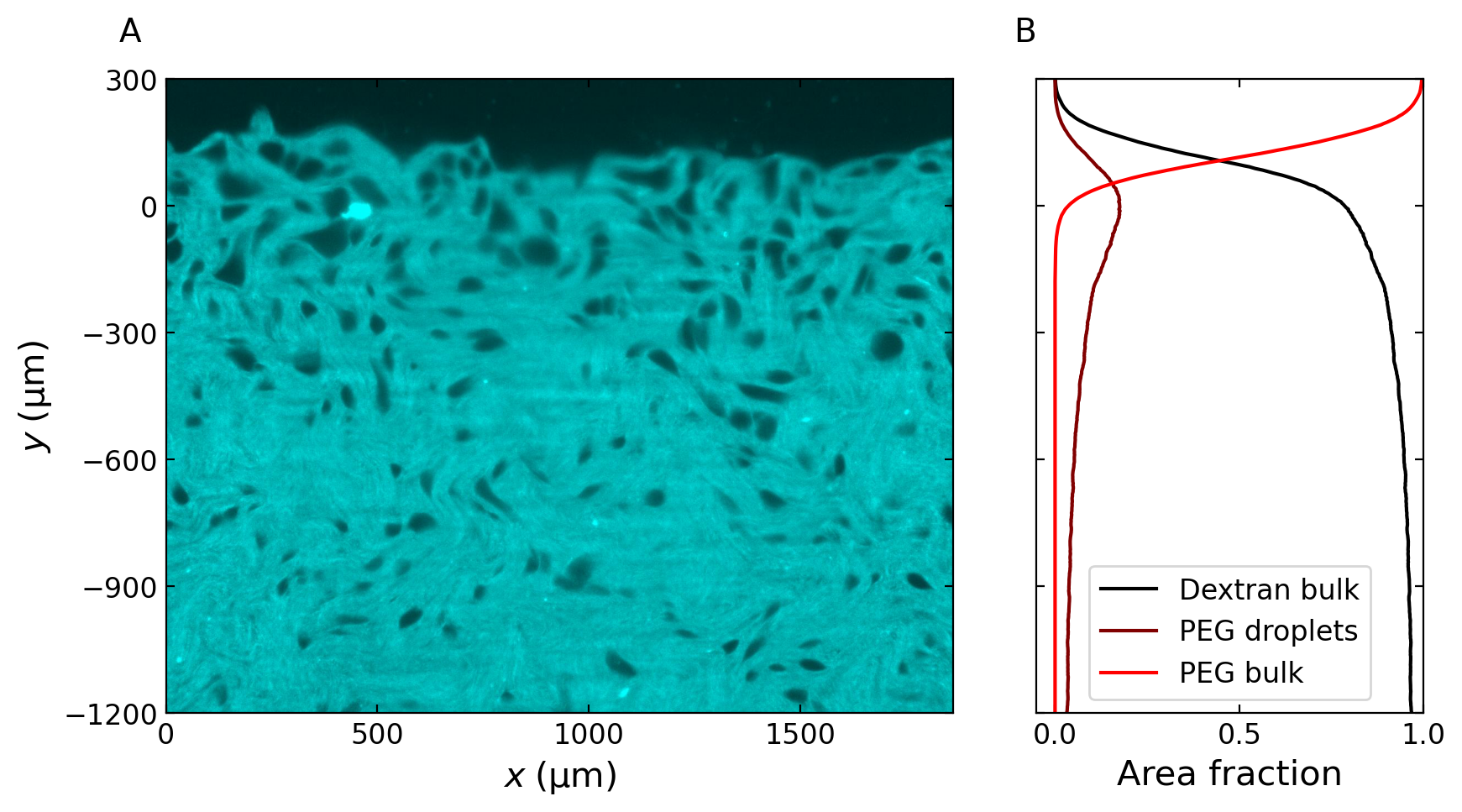}
\end{center}
\caption{The active fluid perforated with passive PEG droplets. (A) Two-dimensional cross-sections of the perforated active phase at $t=2.9$~h. (B) Fractions of bulk dextran phase, PEG droplets, and bulk PEG phase at each depth $y$ in the same frame. The gravitational field points along negative $y$. The sample contained 183~nM KSA, 2.0\% PEG, and 2.0\% dextran.}
\label{droplet}
\end{figure}

We analyzed the structure of the perforated active phase by measuring the position and size of the passive droplets and computing their total volume within the active phase as a function of time (Fig.~\ref{fig:composite_invaginatedArea}C). The total volume was initially zero. It increased rapidly after $t=1.2$~h, when the interface deformations were large enough to generate invaginations. This increase continued  with time, with no hint of saturation over the sample lifetime. When compared to the ``low'' surface tension sample the droplet volume in the ``higher'' surface tension sample increased more slowly and remained smaller. Passive droplets have a lower density than the active fluid. Thus, they creamed towards the PEG/dextran interface creating a perforated foam-like region of the active fluid. Due to gravity mismatch, the density of the droplets varied with the vertical distance from the interface. The fraction of PEG droplets decayed exponentially with increasing distance from the interface (Fig.~\ref{droplet}B).

Using the continuum model, we investigated how increasing activity influences interface behaviors. Beyond a critical value, we observed the envelopment of passive droplets by a pathway similar to experiments (Fig.~\ref{fig:invagination}). To quantify the onset of droplet invagination, we computed the total area fraction of passive fluid that is enveloped by the active phase. For low activities, the total droplet volume is zero. Above a critical value $\alpha_c$, it increases monotonically, saturating at the highest activities (Fig.~\ref{fig:composite_invaginatedArea}D). In simulations, the total volume of enveloped droplets increases as a function of activity. In experiments we obtained a similar behavior but as a function of time. Analogous trends were also observed when analyzing the magnitude of interfacial fluctuations (Fig.~\ref{fig:curvature_exp_sim}). These trends suggest that in experiments the active stress increases with time. The microscopic mechanism for this is not understood. 

\begin{figure}[t]
    \centering    \includegraphics[width=0.49\textwidth]{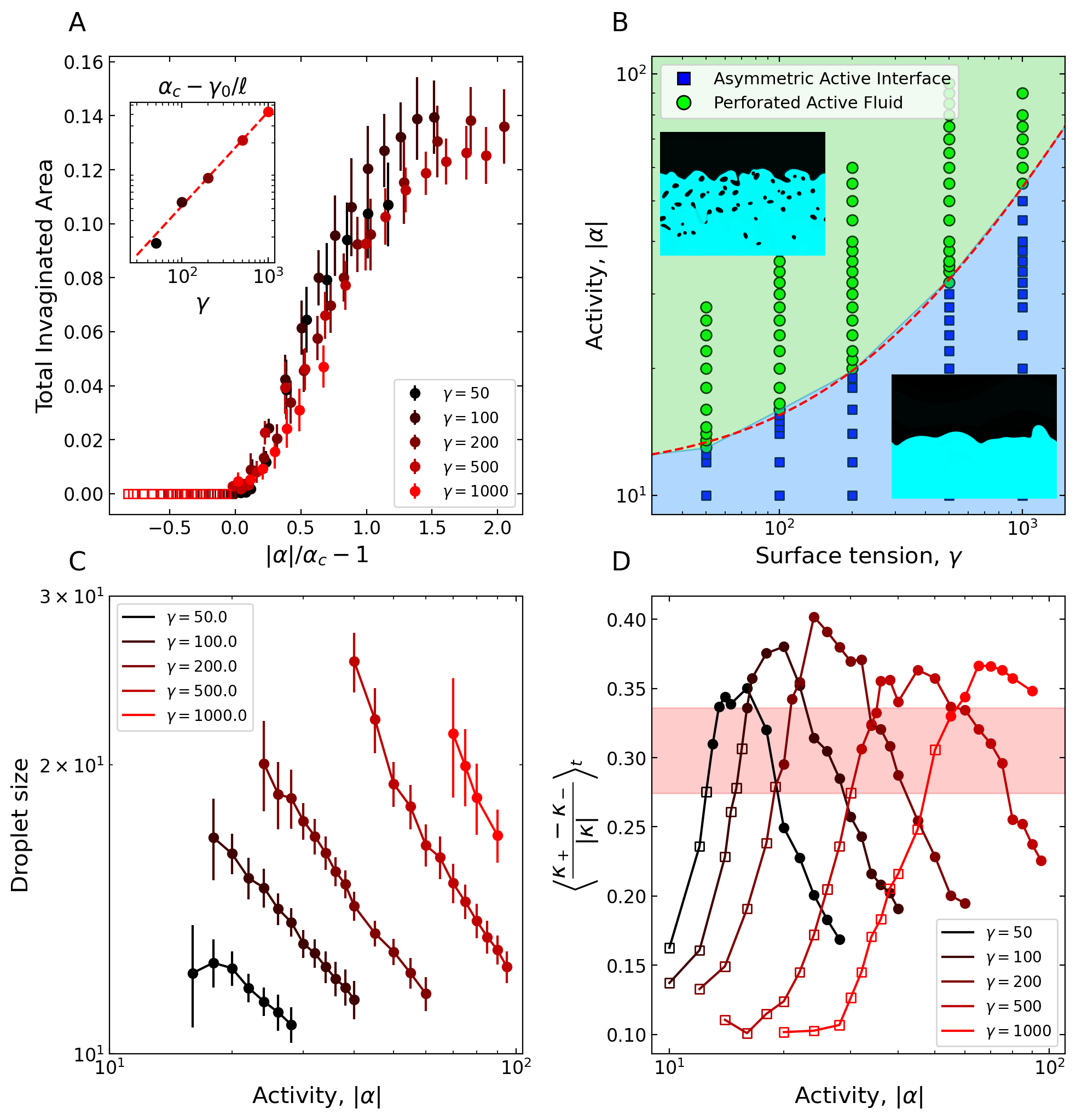}
    \caption{Phase diagram of active interfaces.(A) Total area fraction of passive fluid enveloped by the active fluid as a function of scaled activity $|\alpha|/\alpha_c-1$ for various values of interface tension $\gamma$. The critical activity $\alpha_c$ is numerically approximated as the intercept of the invaginated area with the horizontal axis in Fig. \ref{fig:composite_invaginatedArea}D. Inset: $\alpha_c$ depends linearly on $\gamma$ as $\alpha_c  = (\gamma + \gamma_0)/\ell$. $\gamma_0, \ell$ are fitting parameters. (B) The phase diagram in the activity/tension plane. The dashed red line indicates  the transition from asymmetric interfacial fluctuations to the perforated active fluid.  Insets: images of the phase field $\phi$ at steady state in each phase (asymmetric interface: $\alpha=-18$, $\gamma =200$; perforated state: $\alpha=-28$, $\gamma =200$). (C) The dependent of the droplet mean size on $|\alpha|$ for various interfacial tensions. (D) Normalized mean curvature asymmetry of the interface vs. $|\alpha|$ for various $\gamma$. The region highlighted in pink corresponds to the onset of droplet invagination, which occurs at a critical value of the curvature asymmetry. At large activity, the asymmetry decreases due to droplet invagination which removes the high-curvature regions.}
    \label{fig:phasediagram}
\end{figure}

We use the total droplet volume as an order parameter to quantify the transition to the perforated active state. The interfacial tension controls the location of the transition, with higher tension samples requiring higher activity (Fig.~\ref{fig:composite_invaginatedArea}D). Intriguingly all the plots of droplet volume as a function of activity collapse onto a master curve by rescaling the activity (Fig.~\ref{fig:phasediagram}A). Using these data, we constructed a phase diagram showing the stability of bulk phase separation versus a perforated state as a function of activity and interfacial tension (Fig.~\ref{fig:phasediagram}B). For $|\alpha|<\alpha_c$ there are asymmetric interfacial fluctuations but no droplet formation. For $|\alpha|>\alpha_c$ the active fluid is perforated by passive droplets that are then continuously advected and sheared by the spontaneous flows. The critical activity $\alpha_c$ grows linearly with the interfacial tension: $\alpha_c = (\gamma + \gamma_0)/\ell$, where $\gamma_0$ and $\ell$ as fitting parameters (inset of Fig.~\ref{fig:phasediagram}A).

In the simulations, invaginated droplets can break apart or evaporate through diffusion~\cite{Fernando}. However, they are largely unable to overcome the elastic energy of the intervening liquid crystal and rarely merge with each other or with the bulk passive fluid (Movie S8). At moderate activities, this generates a steady state with a well-defined mean average droplet size (Fig.~\ref{fig:phasediagram}C). At larger activities, however, finite size effects become important as passive droplets are often advected to the bottom of the simulation box where they are replaced by active material due to boundary conditions $\phi(y= -L_y/2) = 1$  and $\phi(y=L_y/2)=0$ (Materials and Methods). This, together with the constant release of interfacial curvature associated with droplet invagination, saturates the curvature asymmetry (Fig.~\ref{fig:curvature_exp_sim}D).

The transition to the perforated state correlates with the normalized asymmetry of the interfacial curvature $(\kappa_+-\kappa_-)/|\kappa|$, where $|\kappa| = \int_{-\infty}^{\infty}d\kappa \rho(\kappa) |\kappa|$ is the mean of the magnitude of the local curvature. The transition to the perforated state occurs when the curvature asymmetry exceeds a threshold value, corresponding to the pink band in Fig.~\ref{fig:phasediagram}D. A minimal condition for droplet invagination is that the active pressure $|\alpha|$ exceeds passive restoring pressure from surface tension and gravity $\sim \gamma/\ell  + \Delta \rho g h$, where $h$ and $\ell$ characterize the size of fluctuations. When these fluctuations grow large enough, passive droplets are enveloped by the active phase in the regions of high negative curvature (Fig.~\ref{fig:invagination}). This decreases the mean curvature asymmetry which keeps decreasing with increasing activity as more and more droplets are pinched off the interface (Fig.~\ref{fig:phasediagram}D).
\section*{Discussion}
We studied the behavior of bulk active interfaces. Our work revealed an up-down asymmetry of the interfacial fluctuations driven by an active fluid. We showed that such symmetry breaking is due to interface-adjacent vortical flows generated by active stresses located only on one interface side. At higher activities, we observed giant interfacial fluctuations, which led to the self-folding of the interface and the formation of passive droplets enveloped by the active fluid. These behaviors were observed in both experiments and numerical simulations. They should be common to all extensile active fluids with liquid crystalline degrees of freedom.

We compare our findings to previous work that studied active interfaces in quasi-2D samples confined between vertical glass walls separated by $60~\mu$m spacers~\cite{ray}. Confined interfaces did not exhibit measurable asymmetries. This difference can be understood using our 2D model. In the confined samples, walls screen the range of active flows~\cite{PoojaPRL}. Incorporating this effect into the numerical model through frictional damping in the Stokes equation yielded no measurable asymmetry of interfacial fluctuations~\cite{ray}. In the 2D model employed here instead, we only include internal viscous dissipation, which yields long-ranged hydrodynamic flows and strong asymmetries of the interface. The qualitative differences between experimental 2D and 3D samples and between 2D simulations with and without friction highlight the key role of long-ranged active in generating asymmetric fluctuations of active interfaces. 

Experimentally we do not have a sufficient time resolution to track passive droplets enveloped by the active phase. Even without such data, however, we can make several qualitative observations. First, the lighter enveloped droplets accumulated near the active-passive interface. Once there they were persistently separated from the bulk passive fluid by a thin layer of active fluid (Movie S5). Thus passive droplets rarely merge back with the bulk passive fluid. We also rarely observed the merger of two droplets. The same behavior was found in numerical simulations. These observations suggest the presence of an effective repulsion between a pair of droplets or a droplet and the bulk interface. 

In the high-activity regime, at late times the total volume of enveloped droplets increases linearly with time at a rate $\approx 1.7\times10^4\mu \mathrm{m}^3/s$ (Fig.~\ref{fig:composite_invaginatedArea}C). Imaging data reveal roughly one invagination event per volume of $4\times 10^4\mu \mathrm{m}^3$ for each frame taken every 4.3 seconds. This allows us to roughly estimate that the flux due to invagination is $\approx 1\times 10^4\mu \mathrm{m}^3/s$, which is comparable to the rate of volume increase. This supports our claim that once invaginated, droplets remain trapped in the bulk active phase. 

\begin{figure}[h]
    \centering
    \includegraphics[width=0.42\textwidth]{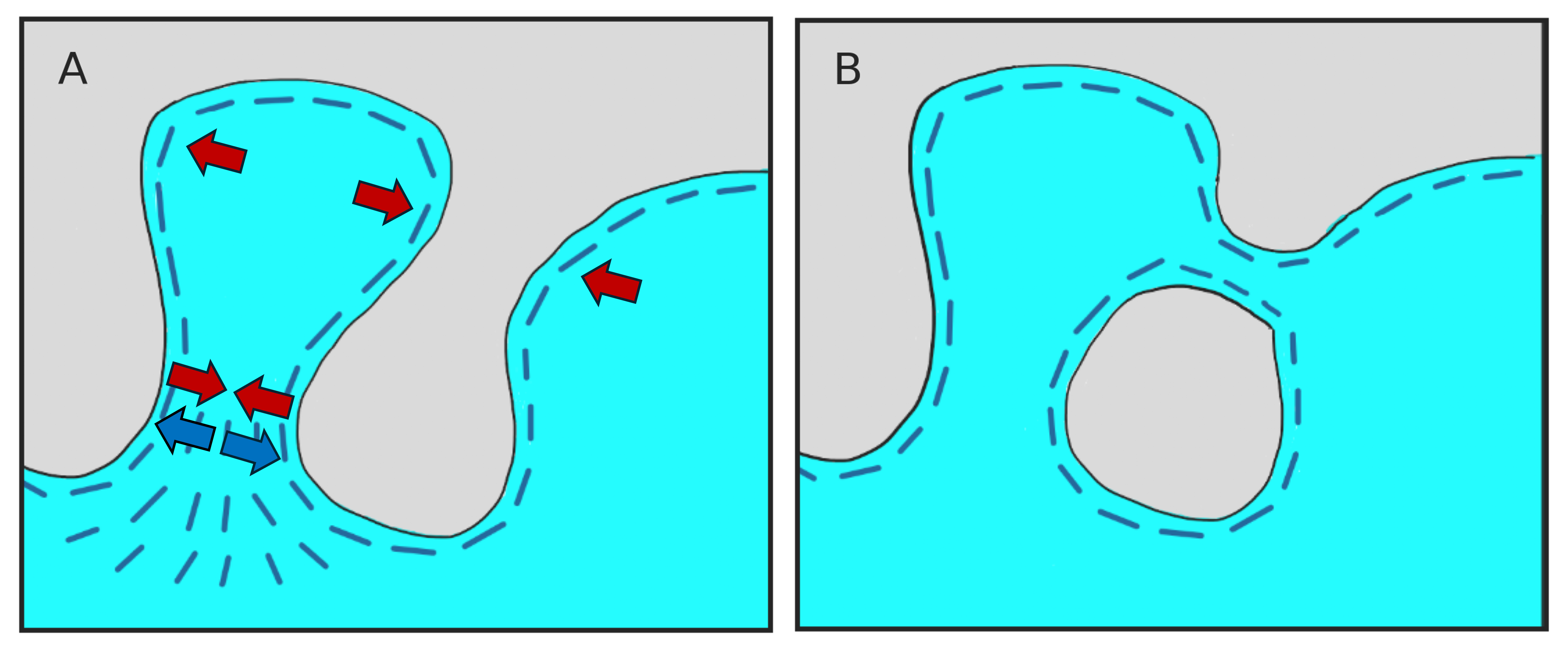}
    \caption{Origin of the asymmetry of droplet splitting at the interface. The dashed line along the interface illustrates active anchoring which yields an aligned interfacial nematic layer. (A) Extensile active forces in the active fluid (red arrows) enhance any curvature of the interfacial nematic layer, promoting the splitting of droplets. At the neck of a protrusion, however, deformations are impeded by elastic forces arising from deformations of the aligned interfacial layer (blue arrows). This leads to an effective repulsion between two interfaces separated by a nematic region. Such repulsions become important when the thickness of the active neck is comparable to the alignment length. They prevent the release of active droplets into the passive phase. The same elastic repulsion prevents passive droplets, once invaginated, from merging back with the bulk passive fluid (Movie S9). (B) The invagination of a passive droplet into the active phase requires the lateral growth of the vertical walls and the reduction of the intervening passive region.  The two vertical walls merge easily, as the intervening passive region can shrink without raising elastic energy, and hence does not impede this process.}
    \label{fig:sketch_elastic_repulsion}
\end{figure}

We hypothesize that repulsion between two passive droplets and passive droplet and bulk interface results from active anchoring, which creates an interface-bound aligned layer of the liquid crystals. When the interface is deformed, the resulting elastic energy maintains the separation between interfaces (Fig.~\ref{fig:sketch_elastic_repulsion}). This elastic repulsion between interfaces separated by active fluid regions becomes important when the separation is comparable to the alignment length, which in turn is of the order of the active length. This effect could also stabilize the perforated active phase, preventing the merging of passive droplets back into the passive bulk fluid.

We elucidated two complementary mechanisms that lead to asymmetry of active interfaces over a broad range of activities. The interface breaks the symmetry of the fundamental bend instability, while nematic elasticity suppresses the merging of passive droplets and the break-up of bulk active phase into finite-sized droplets. These are, however, just partial ingredients required to develop a complete model of deformable active interfaces. For example, in a bulk fluid, the fully developed bend instability leads to local fracture and generation of topological defects~\cite{Sanchez}. The interactions of topological defects with a soft interface is a fertile and relatively unexplored area. Defects deform soft interfaces asymmetrically. In turn asymmetric interfaces can control the spatial nucleation of topological defects and their subsequent motility (Fig.~S7)~\cite{young2021many,xu2023geometrical,chaithanya2024transport,bhattacharyya2024phase,ruske2021morphology}.

We note a broad qualitative agreement between experiments and theory. There are, however, also differences. First, simulations reach a dynamical steady state, as evidenced by the total volume of enveloped droplets that saturates at long times (Fig.~S9). The experimental system, in contrast, shows no sign of reaching a steady state. One reason is that in theory it is easy to fix activity. In comparison, several observations suggest that the activity in microtubule-based fluids increases over time (Fig.~\ref{fig:curvature_exp_sim}B,~\ref{fig:composite_invaginatedArea}C), which could be causing growth of the volume of the passive phase trapped in the active fluid  (Fig.~\ref{fig:composite_invaginatedArea}B). Another  reason for this difference is that in simulations invaginated droplets split into smaller droplets which can then evaporate via an Ostwald ripening-like process. This process  of Oswald-like rippening is not observed in experiments. Another notable difference is that in experiments  the mean size of individual invaginated passive droplets gets larger with time (Fig.~S12), while it decreases with increasing activity in simulations (Fig~\ref{fig:phasediagram}C, S9)

The perforated active phase qualitatively resembles the microphase-separated state predicted in a scalar model of active phase separation~\cite{fausti2021capillary}. There are, however, important differences. In the scalar model, active forces only act at the interface resulting in an effective negative interfacial tension. In our system, in contrast, liquid crystalline degrees of freedom and associated active flows in the bulk liquid crystal play an important role in driving the asymmetry of interfacial fluctuations and droplet invagination. Whether the interface between the active liquid crystal and a passive fluid can be described by a model using an effective single scalar model is an open question.

More broadly, we note intriguing yet poorly explored similarities between active phase separation and phase separation of two immiscible viscoelastic polymers~\cite{tanaka2000viscoelastic}. Upon phase separation of two polymers with very different relaxation times, one often observes the creation of a sponge-like network of the slow component trapped inside the fast phase, with an associated shrinking of the volume of the bulk slow phase~\cite{tanaka1996universality}. This phenomenon arises from the asymmetric distribution of viscoelastic stresses among the two components and it is a long-lived transient, with the long stages of viscoelastic phase separation approaching the behavior of conventional fluid phase separation, as described, for instance, by Model H~\cite{chaikin1995principles}. In the absence of activity, our system consists of two phase separating polymers, PEG and dextran, but their passive phase separation is not affected by viscoelasticity: the volumes of the two bulk phases remain constant during phase separation and no trapping of one phase into the other is observed. In the presence of activity, however, the two components have vastly different dynamics, as active stresses act directly only on the active dextran/MT liquid crystals and accelerate their dynamics relative to that of the passive PEG fluid. The perforated active phase bears qualitative similarities with the sponge-like state observed in viscoelastic phase separation but does not appear to be a transient state. Investigating the connections between active and viscoelastic phase separation is an interesting direction for future work.

Our findings have broad implications. For example, in biology, the stress generating cytoskeleton interacts with membrane-less organelles and lipid membranes, driving their non-equilibrium fluctuations and associated processes such as endo- and exocytosis. Our results have implications for understanding these complex interactions. From a materials science perspective, a self-folding active interface provides a pathway for generating finite-sized droplets and vesicles, by a mechanism that is fundamentally different from the current microfluidic technologies. 
\section*{Materials and Methods}
\textbf{Protein purification.}
Tubulin monomers were purified and labeled by established protocols \cite{Tayar2022}. 3\% of tubulin was labeled with Alexa Fluor 647 NHS Ester (Invitrogen) using a Succinimidyl ester linker. The labeled and unlabeled tubulin were polymerized together into microtubules (MT) with 0.6~mM guanylyl-(alpha,beta)-methylene-diphosphonate (GMPCPP)(Gena Biosciences) and 1.2~mM Dithiothreitol (DTT) in M2B buffer\cite{Tayar2022}. 
The M2B Buffer contained 1~mM ethylene glycol-bis($\beta$-aminoethyl ether)-N,N,N',N'-tetraacetic acid (EGTA), 2 mM MgCl2 and 80~mM piperazine-N,N'-bis(2-ethanesulfonic acid) (PIPES) in water and the pH was set to 7 \cite{Tayar2022}. The final concentrations of MTs in the samples were 0.67~mg/mL. Kinesin-401 protein fused to a biotin-carboxyl carrier domain was expressed and purified by established protocols \cite{Tayar2022}. It was then attached to tetrameric streptavidin in an M2B buffer with 5~mM DTT, forming a kinesin-streptavidin cluster (KSA). The final concentrations of KSA in the samples varied from 42-183~nM. 

\textbf{PEG-dextran active phase separation.}
The active phase separated samples were based on the previously published protocol, with several modifications~\cite{ray}. Previous experiments used a mixture of 35~kDa PEG  and 2~MDa dextran. Activity in these samples could only be sustained by a specific polymer concentration, which did not allow for the control of the surface tension. Therefore, the phase separation here was based on a new protocol. We added PEG (100~kDa, Sigma-Aldrich) and dextran (450-650~kDa, Sigma-Aldrich) into M2B buffer (80~mM K-pipes, 2~mM MgCl$_2$, 1~mM EGTA, pH 6.8). For visualization we added a small amount of amino-dextran (Fina Biosolutions, ($<$ 0.1\% w/w final concentration)) labeled with Alexa-Fluor 488 NHS Ester (Invitrogen). The final sample contained 2.0-2.1\% (w/w) PEG and 2.0-2.1\% (w/w) dextran. 

The MTs and KSA were added to the PEG-dextran mixture with ATP (3~mM final), antioxidants, and an ATP regeneration system made by established protocols \cite{ray}. The antioxidants contained (in final concentrations) 2~mM Trolox, 5~mM DTT, 3.3~mg/mL glucose, 200~$\mu$g/mL glucose oxidase, and 35~$\mu$g/mL catalase dissolved in a phosphate buffer (20~mM K$_2$HPO$_4$ and 100~mM KCl in DI water, pH 7.4). The ATP regeneration system contained (in final concentrations) 26~mM phosphoenol Pyruvate (PEP) (Beantown Chemical) and pyruvate kinase/lactate dehydrogenase (Sigma Aldrich) dissolved in M2B. 

\textbf{Chamber.}
The sample was contained in a transparent fluorinated ethylene propylene(FEP) tube with an inner diameter of 2.4~mm, outer diameter of 4.0~mm, and refractive index of 1.344. The FEP tube was placed into a slightly larger aluminum tube and heated in an oven with ${80}^{\circ} \mathrm{C}$ for 20~min for straightening \cite{FEPtreatment}. The tube was sonicated (Branson 3800) in 1~M KOH for 30~min and 190 proof ethanol for 30~min and stored in 10\% (w/w) Pluronic F-127 (Sigma-Aldrich) solution in water. The tube was sealed from the bottom with UV glue and parafilm. It was mounted to the microscope from the top. 

\textbf{Microscopy.}
The sample was loaded into the FEP tube with a pipette and centrifuged at $2000g$ RCF for 1 min (Fisher Scientific, 05-090-128) to bulk separate the PEG-rich and dextran-rich phases. The tube was mounted onto a laser sheet microscope (Zeiss Z.1 Lightsheet) and imaged for 3 hours. The tube was immersed in water contained in a cubic transparent container to keep the optical path the same during 3D acquisition. The temperature was ${20}^{\circ} \mathrm{C}$. The illumination objectives (5x 0.1 NA) create the laser sheet from both sides of the samples. The imaging objective was 5x 0.16 NA. The illumination lasers were 488~nm for the dextran channel and 638~nm for MT channel. In each time frame, $x-y$ images with step size $\Delta z=20\mu$m were taken from the center of the tube, generating a 3D image stack.

\textbf{3D Curvature.}
To quantify the interface, we extracted the boundary between the two phases from the dextran-channel image, which uses a machine-learning algorithm to classify the image based on intensity thresholding~\cite{ilastik}. The 3D interface was transformed into a mesh form using the software MeshLab \cite{meshlab}. The mean curvature on the interface was defined as the average of the 2D curvatures in the $x-y$ and $y-z$ plane and calculated using discrete exterior calculus \cite{DECCrane,DECLab}.

\textbf{Numerical Simulations.}
The equations for the continuum model (Eqs. \ref{eq:continuum_model}) were solved using self-developed pseudo-spectral solvers \cite{caballero2024cupss, Gulati_activeFolding_cuPSS}. The equations were solved on a two-dimensional grid of size $1024\times 1024,$ with grid spacing fixed to be $\Delta x = \Delta y = \sqrt{2/3}$, small enough to accurately resolve the interface between the active and passive phases (which is the smallest length scale in the system). Time evolution consisted of a simple backward Euler time stepping, with $\Delta t \in [10^{-5}, 10^{-4}]$, simulated for a total of $10^6$-$10^7$ time steps. We consider periodic boundaries in the horizontal ($x$) direction, and in the vertical direction ($y$), we consider no-slip boundaries, with $\phi = 0, 1$ fixed at the top and bottom boundaries respectively, consistent with the gravitational force (i.e heavier active phase on the bottom). The system is initialized to be bulk phase separated, with equal volumes of the active and the passive phases and the active phase at the bottom, $\phi (x, y, t=0) = \Theta\left({- y}\right)$. Refer to the SI for additional details.

\section*{Acknowledgements} This work was primarily supported by the US Department of Energy, Office of Basic Energy Sciences under award number DE-SC0019733. Development and optimization of the two-phase system of active liquid-liquid phase separation was supported by NSF-ISS-2224350. We also acknowledge the use of the biosynthesis facility supported by NSF-MRSEC-2011846 and NRI-MCDB Microscopy Facility supported by the NIH Shared Instrumentation Grant 1S10OD019969-01A1. Use was made of computational facilities purchased with funds from the National Science Foundation (CNS-1725797) and administered by the Center for Scientific Computing (CSC). The CSC is supported by the California NanoSystems Institute and the Materials Research Science and Engineering Center (NSF DMR 2308708).

\bibliography{main.bib}

\clearpage
\widetext

\newgeometry{total={6in,8.5in}}

\section{Supplementary material}

\setcounter{figure}{0}
\renewcommand\thefigure{S\arabic{figure}} 

\setcounter{equation}{0}
\renewcommand\theequation{S\arabic{equation}} 

\section{Continuum Model}
The active/passive mixture is described by a phase field $\phi$ representing the conserved concentration of the active phase, the nematic order parameter $\mathbf{Q}$, quantifying the liquid crystalline orientational order present in the active fluid, and the flow velocity $\mathbf{v}$. The dynamics is governed by the following equations,
\begin{align}
    D_t \phi &= M \nabla^2\mu\;,\\
    D_t \mathbf{Q} &= \lambda \mathbf{A} - \boldsymbol{\omega} \cdot \mathbf{Q} + \mathbf{Q}\cdot \boldsymbol{\omega} + \frac{1}{\Gamma}\mathbf{H},\label{eq:full_Q}\;,\\
    0 &= \eta \nabla^2\mathbf{v} +  {\boldsymbol\nabla \cdot \left(\boldsymbol{\sigma}^e +\boldsymbol{\sigma}^\phi+\boldsymbol{\sigma}^a\right)}  - {\boldsymbol\nabla} P + \mathbf{f}_g\;, \label{eq:StokesFlow}
\end{align}
where $A_{ij} = (\partial_i v_j -\partial_j v_i)/2$ and $\omega_{ij} = (\partial_i v_j - \partial_j v_i)/2$ are, respectively, the strain rate and the vorticity, and $D_t=\partial_t+\mathbf{v}\cdot\bm\nabla$ is the advective derivative.
The flow is assumed to be incompressible, i.e, $\nabla \cdot \mathbf{v} =0.$ The chemical potential $\mu =\delta F_\phi/\delta \phi$ is obtained from the 
the conventional Cahn-Hilliard free energy, 

\begin{align}
    F_\phi = \frac{4\gamma}{\xi}\int d\mathbf{r}\,\bigg( \phi^2(\phi- 1)^2 +  \frac{\xi^2}{2}(\nabla\phi)^2\bigg)\;,
    \label{eq:Fphi}
\end{align}
that describes phase separation between an active fluid ($\phi=1$) and a passive fluid ($\phi=0$).
 
The Cahn-Hilliard free energy is written here in terms of the interfacial thickness $\xi$ and the interfacial tension $\gamma$. The interface stiffness  which controls the capillary stress driving passive flows can then be written as $k = \gamma \xi$. 

The molecular field $H_{ij} = - \delta F_Q/\delta Q_{ij}$ is determined by the liquid crystalline free energy, given by 

\begin{align}
 F_Q = \int d\mathbf{r}\,\bigg( \frac{r \phi }{2}  {\Tr\mathbf{Q}^2}  + \frac{u}{4} \left(\Tr\mathbf{Q}^2\right)^2 + \frac{K}{2} ({\partial_kQ_{ij}})^2\bigg)\;,\label{eq:FQ}
\end{align}
where $K$ is the nematic elastic modulus (in a one-elastic constant approximation) and $r$ and $u>0$ are microscopic energy densities that  control the liquid crystalline ground state. 
To model the experimental system, where the density of microtubules is too low to yield bulk nematic order in the absence of activity, we place ourselves in the isotropic state of the active liquid crystals by choosing $r>0$. It is known, however, that even in this regime, active flows induce local nematic order through flow alignment controlled by the parameter $\lambda$ in \eqref{eq:full_Q}. 

The flow is governed by the Stokes equation that balances viscous dissipation with viscosity $\eta$ and gradients of pressure $P$ with gradients of capillary ($\bm\sigma^\phi$), elastic ($\bm\sigma^e$) and active ($\bm\sigma^a$) stresses, given by
\begin{align}
   \sigma^\phi_{ij} &= -k \left(\nabla_i \phi \nabla_j \phi - \frac12 \delta_{ij} (\bm\nabla \phi)^2\right) \;,
   \label{eq:sigma_phi}\\
   \sigma^e_{ij} &= -\lambda H_{ij} + Q_{ik} H_{kj} - H_{ik}Q_{kj}\;,
    \label{eq:sigma_pe}\\
   \sigma^a_{ij} &= \alpha \phi Q_{ij}\;.
    \label{eq:sigma_a}
\end{align}
In addition, the fluid experiences a gravitational force density $\mathbf{f}_g = {-\hat{y}} \rho(\phi) g$, where $\rho(\phi) = \rho_0 + \phi \Delta \rho$ is the density of the fluid mixture, with $\rho(\phi=0)=\rho_0$ the density of the passive fluid  and $\rho(\phi=1)=\rho_0 + \Delta \rho$ the density of the active fluid. Since the active fluid is denser than the passive one, $\Delta \rho >0$. 

The strength of the active stress is controlled by the activity  $\alpha <0$ to model the extensile active forces present in the microtubule-kinesin system. The active stress vanishes in the passive regions where $\phi=0$. In a bulk nematic state any finite activity destabilizes the ordered state. The isotropic state of an extensile active liquid crystal is similarly linearly unstable, but only for $|\alpha|>\alpha_0$, with $\alpha_{0} = 2\eta r/\lambda \Gamma$. In our work, we consider activities much larger than $\alpha_{0}$, which create self-sustained chaotic flows within the active phase.

The above equations contain a number of parameters summarized in Table~\ref{Table:parameters}. We choose the interface width $\xi$ as unit of length, the nematic reorientation time $\Gamma/r$ as unit of time and $r$ as unit of energy density (or pressure in $2d$). In the following we mainly vary activity $\alpha$ and surface tension $\gamma$. Unless otherwise specified, all other parameters are kept fixed to the values given in Table~\ref{Table:parameters}.

\begin{table}[h!]
\centering
\begin{tabularx}{0.98\textwidth} {|
  >{\centering\arraybackslash}X |
  >{\centering\arraybackslash}X |
  >{\centering\arraybackslash}X |
  >{\centering\arraybackslash}X |}
 \hline
 Parameter & Value(s) & Dimensions & Description \\
 \hline
 $\xi$ & $1$ & $[L]$ & Interface width\\
 $\gamma$ & $50.0 - 1000.0$ & $[E][L]^{-1}$ & Surface tension\\
 $M$ & $0.67$ & $[E]^{-1} [T]^{-1} [L]^4$ & Mobility\\
 $r$ & $1$ & $[E] [L]^{-2}$ & LdG parameter\\
 $u$ & $10.0$ & $[E] [L]^{-2}$ & LdG parameter\\
 $K$ & $106.67$ & $[E]$ & Nematic Elastic Modulus \\
 $\Gamma$ & $1$ & $[E] [T][L]^{-2}$ & Inverse mobility \\
 $\Delta \rho g$ & $0.12$ & $[E] [L]^{-3}$ & Weight difference \\
 $\eta$ & $1.0$ & $[E] [T][L]^{-2}$ & Viscosity \\
 $-\alpha$ & $10.0 - 100.0$ & $[E][L]^{-2}$ & Activity\\
 $\lambda$ & $1.0$ & --- & Flow alignment\\
 $\Delta x = \Delta y$ & $0.82$ & $[L]$ & Lattice spacing \\
 $\Delta t$ & $10^{-5} - 10^{-4}$ & $[T]$ & Simulation time-step \\
 \hline
\end{tabularx}
\vspace{10pt} 
\caption{List of parameters used in the simulations: lengths are measured in units of the interface width $\xi$, times in the units of the nematic reorientation time $\Gamma/r$, and stresses in units of the nematic condensation energy density $r$.}
\label{Table:parameters} 
\end{table}

We integrate the equations numerically using self developed pseudo spectral solvers on a discrete two-dimensional grid \cite{caballero2024cupss, Gulati_activeFolding_cuPSS}. We use pseudospectral solvers since Fourier methods are generally more numerically stable when dealing with higher order gradients of the continuum fields such as the stiffness term required in the Cahn-Hilliard equation $\sim k \nabla^4 \phi$. The integration in time is done with a simple backwards Euler stepping, with a fixed time-step. The lattice size is fixed to  $N\times N =1024 \times 1024$, and we typically run our simulations for $10^6 - 10^7$ time steps.

\begin{figure}[h!]
    \centering
\includegraphics[width=0.35\textwidth]{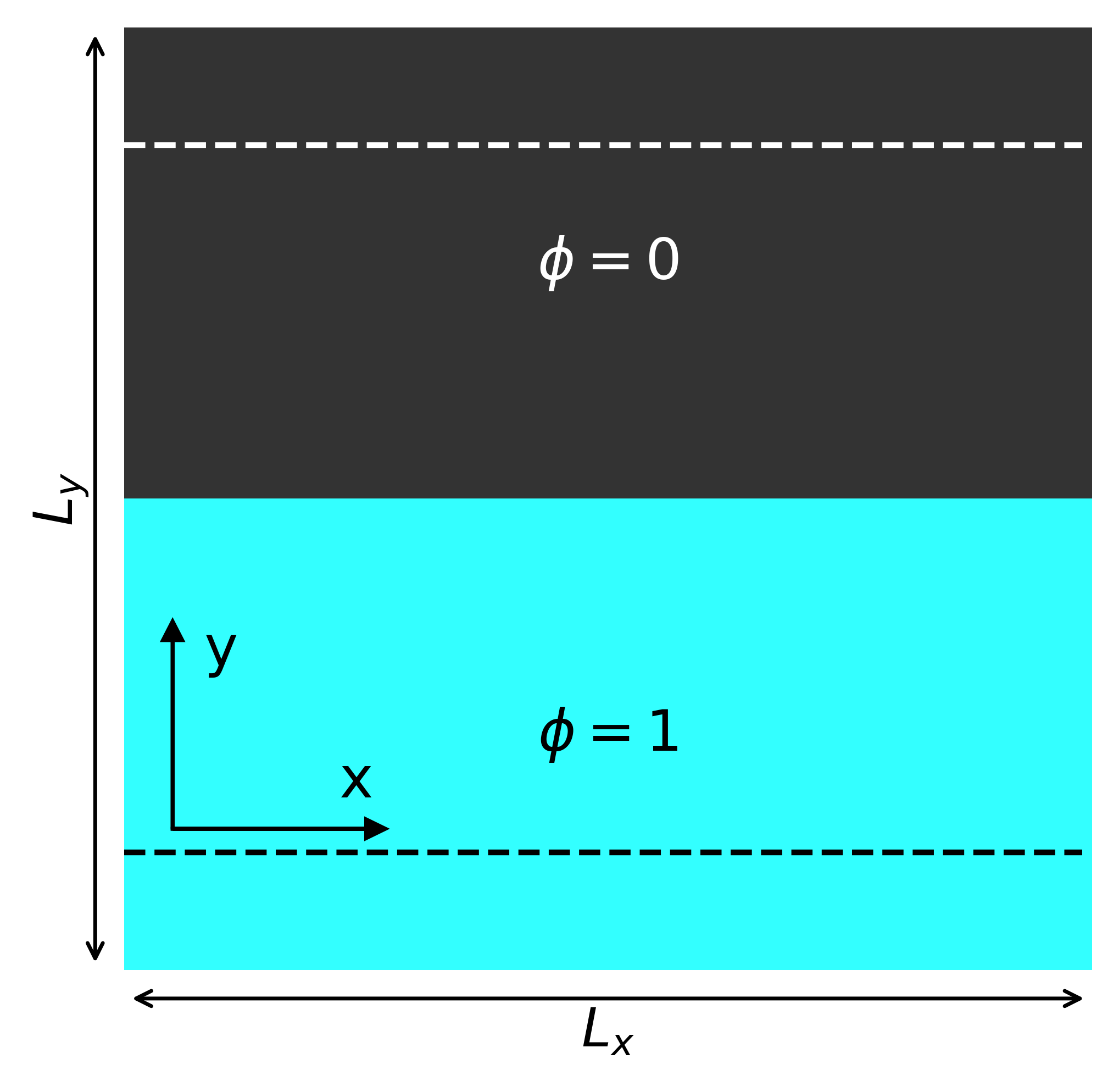}
    \caption{Initial configuration used in the simulations.
    We initialize the system in the bulk phase separated state with no nematic order nor flow. The dashed lines delimit the boundary layers used to impose Dirichlet boundary conditions at $y=\pm  L_y/2$ in the spectral solver. 
    We use  periodic boundary conditions in the $x$ direction. All the simulation results shown are for fixed system size $L_x=N\Delta x$ and $L_y=N\Delta y$, with $N =1024$. 
    }
    \label{fig:sketch_geometry}
\end{figure}

We use periodic boundary conditions in the horizontal ($x$) direction (see Fig.~\ref{fig:sketch_geometry}). We enforce Dirichlet boundary conditions at the top and bottom boundaries by maintaining a band of width $N/8$ at the bottom of the system where $\phi = 1$, representing an active fluid, and a corresponding band where $\phi=0$ at the  top boundary, enforcing a passive fluid. This is consistent with the effect of gravity that segregates the heavier active phase to the bottom of the sample. We also impose no-slip boundary condition at the top and bottom of the sample by requiring $\mathbf{v} = 0$. We initialize the system with a bulk phase separated configuration and equal volumes of the active and the passive phases, with the heavier active phase at the bottom, i.e., $\phi (y,t=0) = \Theta(- y)$, where $\Theta$ is the Heaviside step function. We additionally initialize the system with no nematic order, i.e., $\mathbf{Q}(\mathbf{r},t=0) = 0$ and add a small initial white noise to the nematic field to perturb the system.

\section{Active Anchoring}
To quantify the degree of interfacial anchoring in the active phase, we define the local alignment field $A(\mathbf{r})$ that measures the alignment between the bulk interface normal ($\hat{m}$) and the local nematic director ($\hat{n}$). 
\begin{figure}[h!]
    \centering
\includegraphics[width=0.5\textwidth]{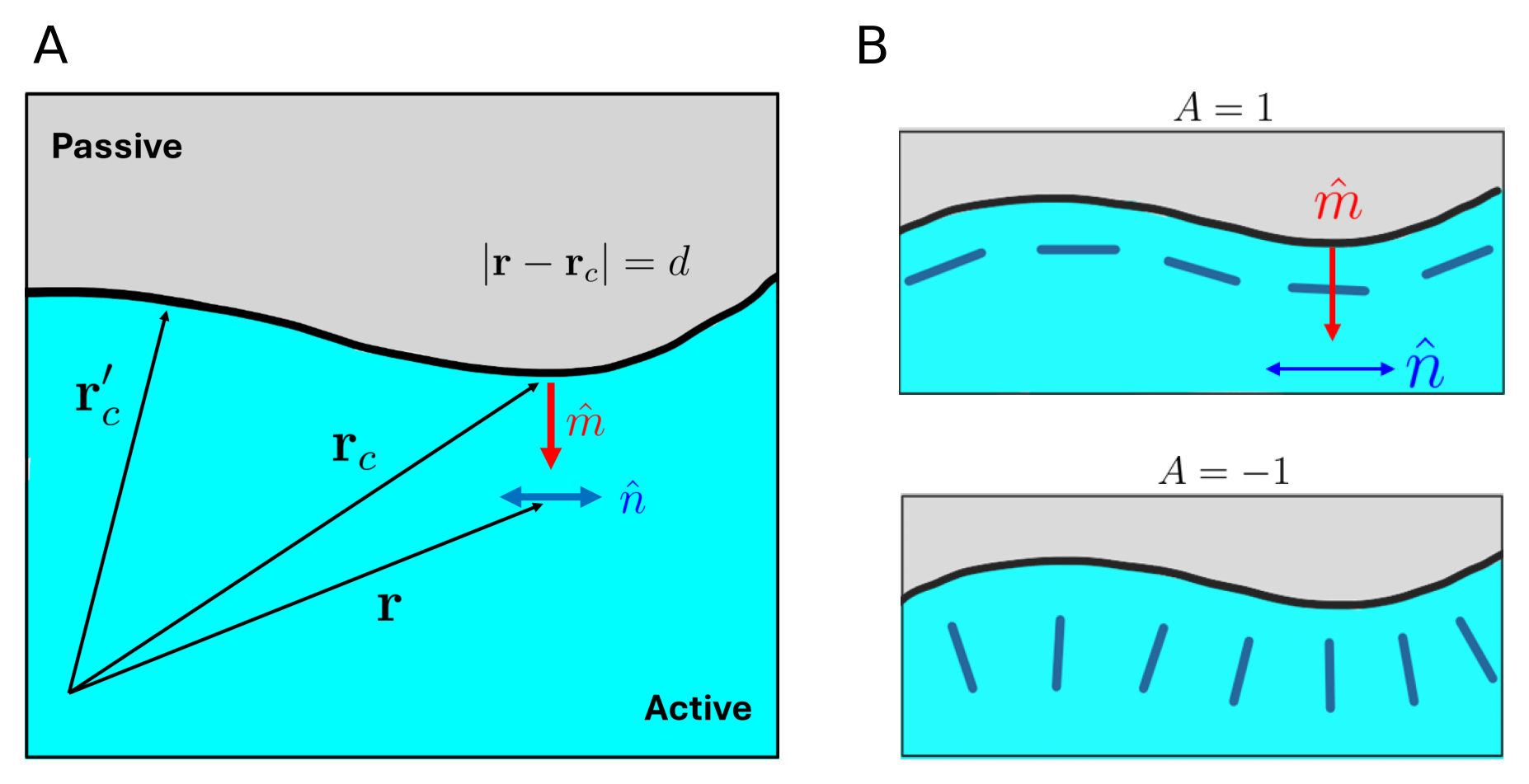}
    \caption{(A) $A(\mathbf{r})$ measures the alignment between the director $\hat{n}$ at $\mathbf{r}$ and the interface normal $\hat{m}$ at $\mathbf{r}_c$, where $\mathbf{r}_c$ is the point closest to $\mathbf{r}$ on the interface i.e. $d=|\mathbf{r} - \mathbf{r}_c| < |\mathbf{r} - \mathbf{r}_c'| \; \forall \; \mathbf{r}_c' \ne \mathbf{r}_c.$ (B) $A =1 (A=-1)$ corresponds to parallel (normal) alignment of the director with the interface.}
    \label{fig:activeanchoring_coordinates}
\end{figure}

\begin{equation}
A(\mathbf{r}) =  1 - 2 (\hat{n}(\mathbf{r}) \cdot \hat{m}(\mathbf{r}_c))^2\;,    \end{equation}
where $\mathbf{r}$ is any point in the active phase and $\mathbf{r}_c$ refers to the point closest to $\mathbf{r}$ on the interface. If the director is everywhere tangent (normal) to the interface, then $A=1 (A=-1)$ (shown as a sketch in Fig.~\ref{fig:activeanchoring_coordinates}).

\begin{figure}[h!]
\centering
\includegraphics[width=0.70\textwidth]{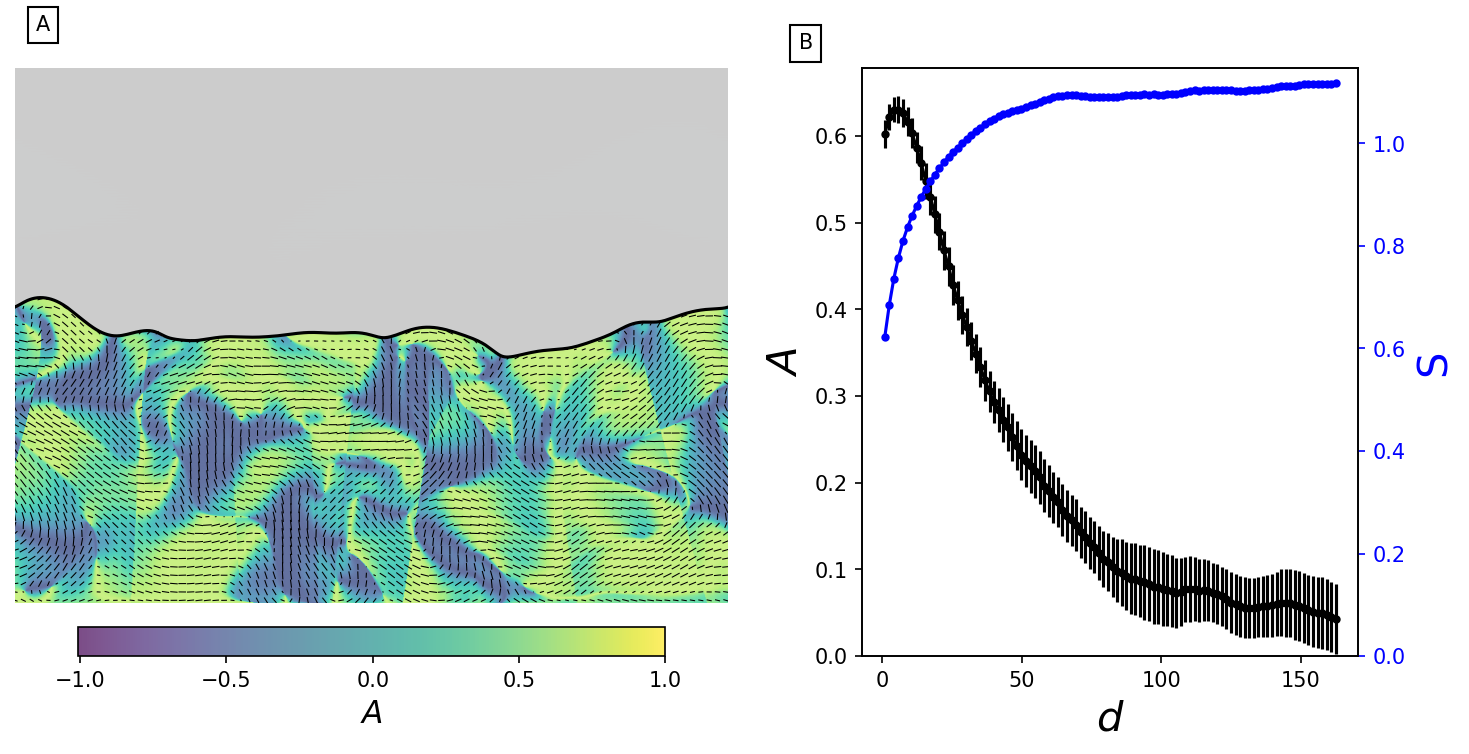}
\caption{(A) Alignment between the director and the interface normal computed in the active phase, $A(\mathbf{r}) =  1 - 2 (\hat{n}(\mathbf{r}) \cdot \hat{m}(\mathbf{s}))^2$. (B) The alignment function and the scalar order parameter $S =2\sqrt{Q_{xx}^2 + Q_{xy}^2}$ are computed as a function of the distance from the interface, $d$, averaged over the steady state.}
\label{fig:activealignment}
\end{figure}

We then compute the mean alignment as a function of the distance $d = |\mathbf{r} - \mathbf{r}_c|$ from the interface, averaging over points distance $d$ from the interface and over time. The amount of alignment with the interface depends on the magnitude of the local nematic order present close to the interface, as quantified by the mean scalar order parameter $S(d)$  at distance $d$ from the interface (Fig \ref{fig:activealignment}).

\begin{figure}[h!]
\centering
\includegraphics[width=0.55\textwidth]{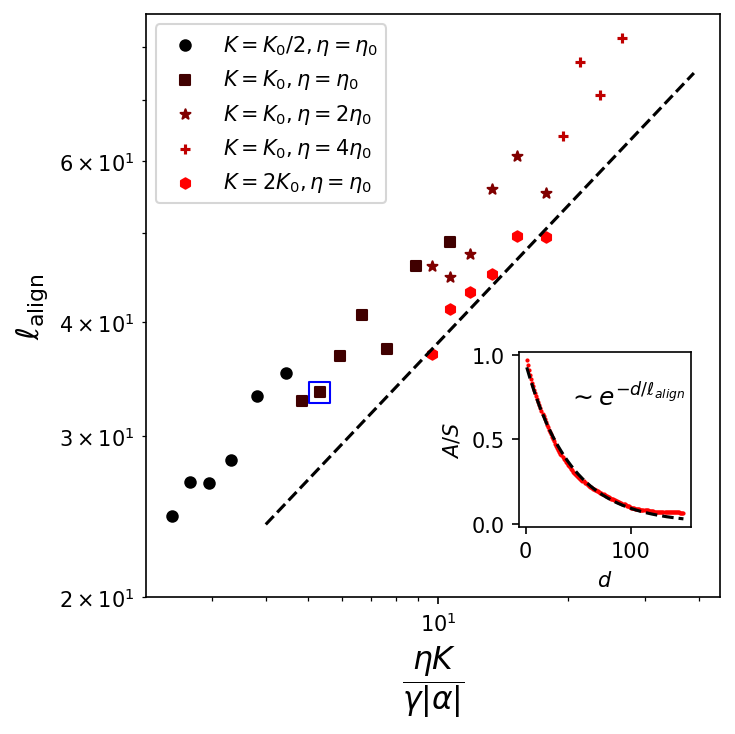}
\caption{We calculate the alignment length from ${A(d)}/{S(d)} \sim e^{-d/\ell_{\text{align}}}$ (shown in the inset), with values of activity $10\ge |\alpha|\le 24$ for various fixed values of the elastic modulus and the viscosity parameters as shown in the legend (at fixed surface tension $\gamma =200$), where $K_0 =106.67, \eta_0=1.0$ are the default values listed in \ref{Table:parameters}. We find $\ell_{\text{align}} \sim \ell_{\text{active}} = \sqrt{\dfrac{\eta K}{\Gamma |\alpha|}}$ (the reference dashed line has slope $=1/2$). The inset corresponds to the point highlighted in blue ($K =K_0, \eta=\eta_0, \alpha =-20$), with the data shown in red, and the exponential fit with a dashed line.}
\label{fig:alignmentlength}
\end{figure}

Indeed, we find that $A(d)/S(d)$ decays exponentially with  $d$. We compute the alignment decay length $\ell_{\text{align}}$ by fitting $A(d)/S(d) = C \exp(-d/\ell_{\text{align}})$. We find that $\ell_{\text{align}}$ is proportional to the active length $\ell_{\text{active}} \sim \sqrt{\eta K/\Gamma |\alpha|}$
(Fig~\ref{fig:alignmentlength}).

\section{Interface Curvature and Asymmetry}
\begin{figure}[h!]
    \centering
\includegraphics[width=0.85\textwidth]{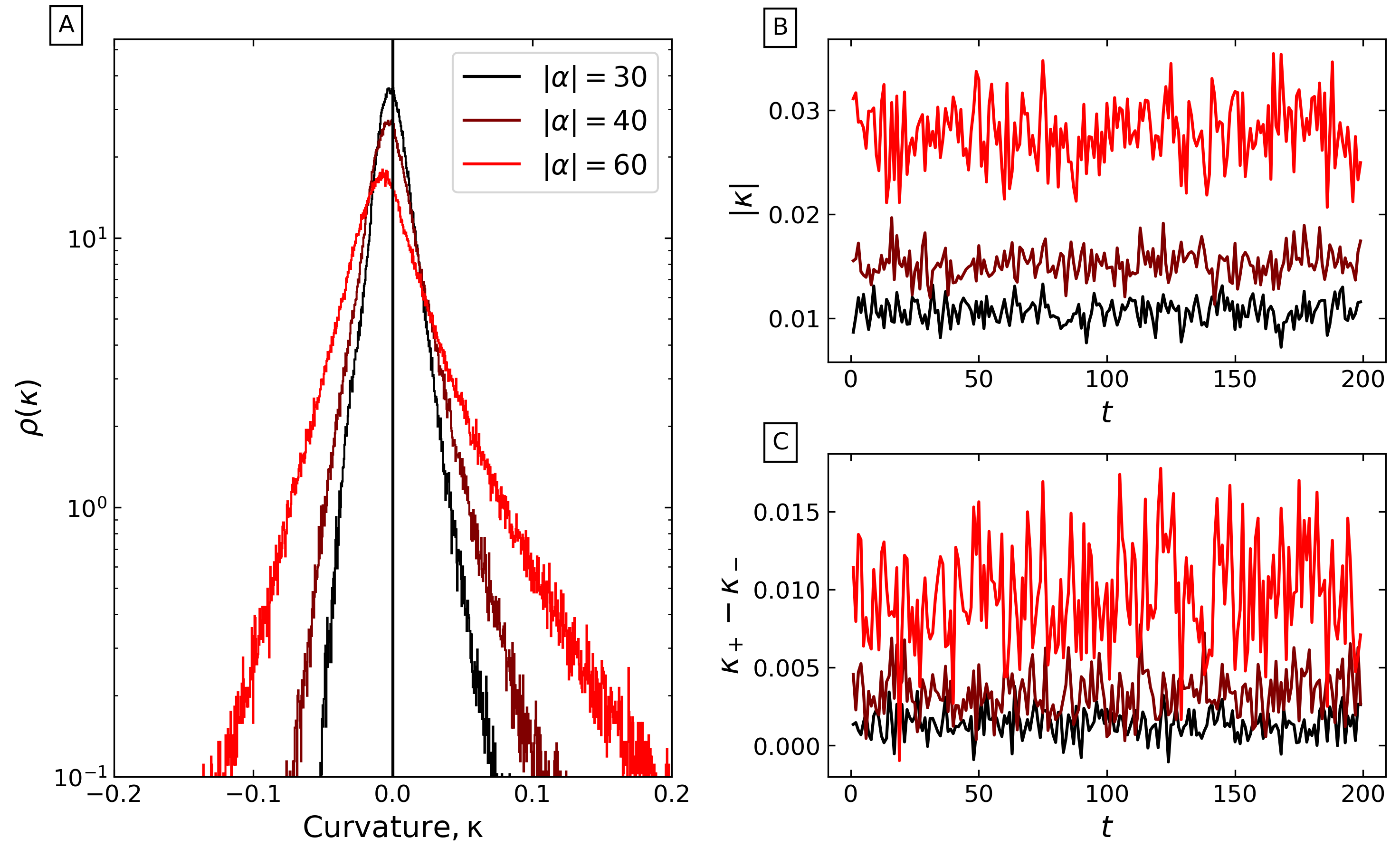}
    \caption{(A) The distribution of the local curvature across the interface for the simulation at a fixed surface tension $\gamma =500$ (same as Fig.~3C). (B, C) The mean (across the interface) of the absolute value of the local curvature, $|\kappa|$ and the asymmetry $\kappa_+ - \kappa_-$ as a function of time.}
    \label{fig:curvature_time_series}
\end{figure}

We define the interface separating the two bulk fluids as the locus of points $(x(s), y(s))$ where $\phi=1/2$. The local curvature of the interface is then given by 
\begin{equation}
    \kappa(s) = \dfrac{x'y'' - y'x''}{(x'^2 + y'^2)^{3/2}}\;.
\end{equation}
where the prime denote derivative with respect to the arclength $s$. As described in the main text, we consider the distribution of these local curvatures across the interface and measure the asymmetry in terms of the positive and negative means of the distribution. Additionally, in Fig.~\ref{fig:curvature_time_series} we show the time evolution of the absolute value of the local curvatures (Fig.~\ref{fig:curvature_time_series}B) and of the asymmetry $\kappa_+-\kappa_-$ (Fig.~\ref{fig:curvature_time_series}C, see Eq.~2 of the main text for definition) to demonstrate that the system has reached a dynamical steady state.

\subsection{Mechanism for the interface asymmetry}

As discussed in the main text, the curvature asymmetry of the interfacial fluctuations can be understood as a generic consequence of the fact that the interface breaks the spatial symmetry of the bulk instability. 
To demonstrate this, we consider the simplified case of an ordered nematic liquid crystal and assume the director field is always tangent to the interface. We then compare the active forces generated by a sinusoidal deformation of the interface between the active and passive phases to a similar deformation of the  director field in a bulk active nematic.     

Consider a phase separated mixture of an active nematic and a passive fluid  described by the phase field $\phi(x, y) = \Theta(h(x) - y)$, with an interface profile of the form
\begin{equation}
    h(x) = A \sin(2\pi x/L_x)\,,
\end{equation}
where $A$ is the amplitude  of the interface fluctuation. Since we assume perfect alignment of the nematic with this interface, the director is given by
\begin{align}
    n_x = \dfrac{1}{\sqrt{1+ (\partial_x h)^2}}, \quad n_y = \dfrac{\partial_x h}{\sqrt{1+ (\partial_x h)^2}}\;.
\end{align}

The active force due to the nematic texture can be written as the sum of contributions from the gradient of the scalar order parameter (or $\phi$) from the active to the passive phase and from the distortions of the nematic within the active phase, i.e., 
\begin{align}  \mathbf{f}^{\text{active}}= \alpha \bm{\nabla} \cdot \left(\phi \mathbf{Q} \right)= \alpha \mathbf{Q} \cdot \bm{\nabla} \phi + \alpha \phi \bm{\nabla} \cdot \mathbf{Q},
\end{align}
The interfacial contribution due to gradients in the magnitude of the order parameter is then always normal to the interface and acts even on a flat interface (which, in an incompressible fluid, corresponds to a pressure jump across the interface). So, we only consider the contributions of the gradients of the director field. In the active-passive mixture, with the interface described by $h(x)$ as earlier, the active force relevant here is given by
\begin{align}
    f^{\text{active, mix}}_i &=  \Theta(h(x) - y) \left( \alpha \nabla_j \left( n_i n_j -\frac{\delta_{ij}}{2}\right) \right)\;.
\end{align}

In Fig.~4B we plot 
this active force due to the distortions in the director field,
which clearly is finite only in the region below the interface.
In contrast, for a bulk active nematic with the same director profile (Fig.~4A), the active force is given by
\begin{align}
    f^{\text{active, bulk}}_i = \alpha \nabla_j \left( n_i n_j -\frac{\delta_{ij}}{2}\right)\;
\end{align}
and clearly extends throughout the fluid.

Using these expressions for the active force, we  then solve numerically the incompressbile ($\nabla\cdot \mathbf{v} =0$) Stokes' equation with viscous dissipation,
\begin{equation}
\eta\nabla^2 \mathbf{v} - \nabla P + f^{\text{active}} =0
\end{equation}
The resulting flows are also displayed in Fig. 4.

When the active forces are present both above and below the interface, we get the familiar flows corresponding to the bend instabliity for a bulk active nematic (Fig~4C). In contrast, in the case of the fluid mixture, the absence of active forces above the interface creates vortical flows that push together the valleys and pull apart the peaks  (Fig.~4D).

In this argument we assume an ordered nematic state for the liquid crystal phase and perfect alignment of the director with the fluid interface. In the full continuum model, both liquid crystalline order and director alignment at the interface are emergent properties due to activity.

\subsection{Minimal model with explicit anchoring}
To illustrate the combined role of anchoring and active flows in controlling the interface curvature asymmetry, we consider a minimal model of a phase-separated active/passive mixture where the active liquid crystal is in the ordered nematic state and the director field is anchored to the interface via anchoring free energy,  

\begin{align}
    F_{\text{anch}} = -\frac{\chi}{2}\int d\mathbf{r}~ \phi (\nabla_i\phi)(\nabla_j \phi)Q_{ij}\;.
    \label{eq:fanch}
\end{align} 
In this minimal model, the evolution of the $Q$-tensor is determined solely by the free energy, with
\begin{align}
    \partial_t \mathbf{Q} = -\frac{1}{\gamma}\frac{\delta(F_Q+F_{\text{anch}})}{\delta\mathbf{Q}}\;,
\end{align} 
and the incompressible flow is governed by force balance in the Stokes limit,
\begin{equation}
    0 = \eta \nabla^2\mathbf{v} +  {\boldsymbol\nabla \cdot \left(\boldsymbol{\sigma}^\phi+\boldsymbol{\sigma}^a\right)}  - {\boldsymbol\nabla} P\;.    
\end{equation}
Here we include the capillary forces at the interface, but neglect gravity.

We compare two situations: (i) a mixture of an active and a passive fluid, and (ii) a mixture of two identical active fluids. In other words, in (i) the active fluid is present only below the interface, while in (ii) the active fluid is present on both sides of the interface.  In the case of the active-passive mixture (i), the nematic free energy $F_Q$ and the active stress $\sigma^a$ are given as before (Eqs.~\ref{eq:FQ}, \ref{eq:sigma_a}). However, when both fluids are active (ii), $F_Q$ and $\sigma^a$ are independent of $\phi$ i.e. 
\begin{equation}
 F_Q^{\text{active-active}} = \int d\mathbf{r}\,\bigg( \frac{r}{2}  {\Tr\mathbf{Q}^2}  + \frac{u}{4} \left(\Tr\mathbf{Q}^2\right)^2 + \frac{K}{2} ({\partial_kQ_{ij}})^2\bigg)\;,
\end{equation}
\begin{equation}
\bm{\sigma}^{a, \text{active-active}} = \alpha \mathbf{Q}.
\end{equation}
In this case, the nematic interacts with interface only through the anchoring term $F_{\text{anch}}$ and the flow due to the capillary stress.

In both cases we consider an ordered nematic state (by setting 
$-r = u =1.0$) and incorporate director alignment parallel to the interface through the anchoring energy with $\chi <0$. 

We consider the evolution of the interface 
from an initial sinusoidal deformation, corresponding to
$\phi(x,t=0) = \Theta(h(x,t=0) - y)$, with $h(x,t=0) = A \sin(2\pi x/ L_x )$.

In the active/active case,  
the initial sinusoidal perturbation grows symmetrically over time, developing no curvature asymmetry (Fig.~5A). In the active/passive mixture, however, we can clearly see the effect of the vortical flows close to the interface which, over time, leads to the formation of sharp valleys into the active phase (Fig.~5B). 

We use this setup only as an illustrative example to demonstrate that asymmetric fluctuations are simply the result of the lack of active material on one side of the interface. 
Since we do not include restoring forces due to gravity, the bulk phase separated state is unstable in general and the interface breaks up over time.

\subsection{Curvature asymmetry and fluctuation amplitude}
\begin{figure}[h!]
\centering
\includegraphics[width=0.7\textwidth]{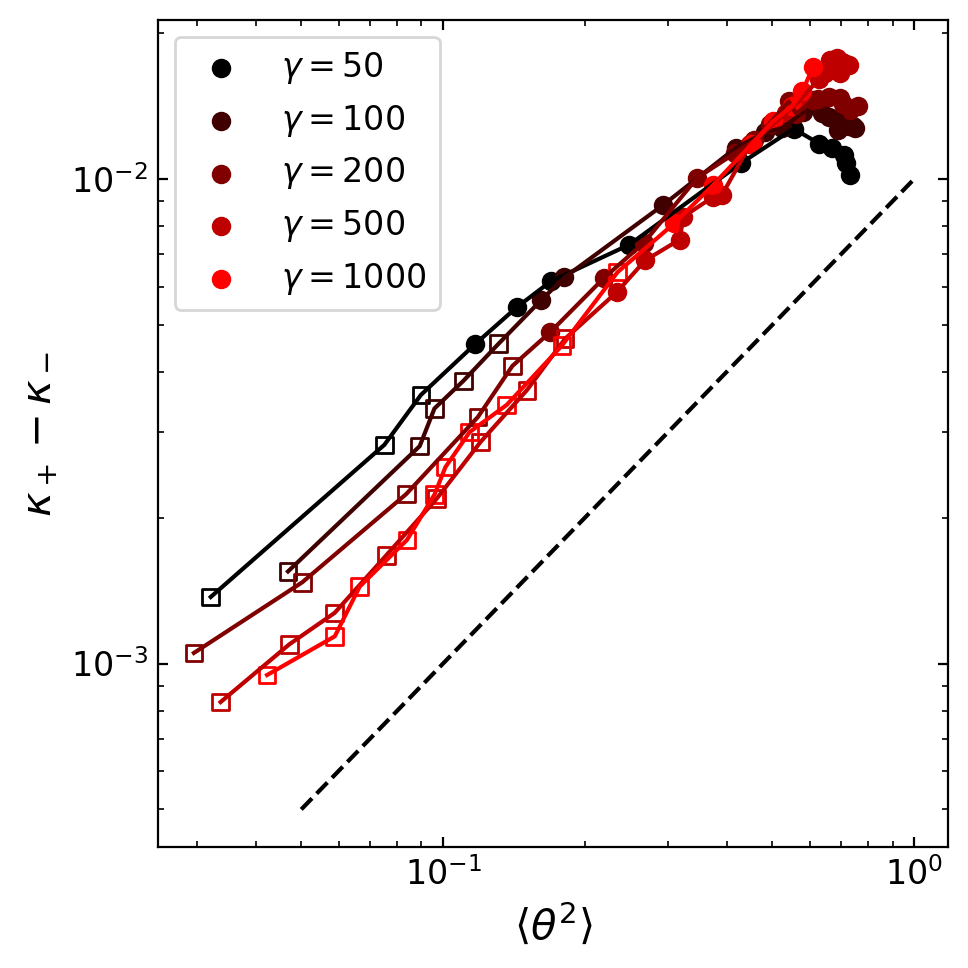}
\caption{We plot the mean curvature asymmetry $\kappa_+ - \kappa_-$ versus the mean amplitude of the interface fluctuations $\langle \theta^2 \rangle$ in steady state. The different data points correspond to the different values of activity shown in the phase diagram (Fig~9). The reference dashed line has slope = 1.}
\label{fig:amplitude_asymmetry}
\end{figure}

In the main text, we have quantified how the curvature asymmetry increases with activity and is suppressed by increasing interfacial tension. Active flows drive both the amplitude of the fluctuations as well as the curvature asymmetry, and the balance between this active driving and restoring forces due to interface tension and gravity creates a dynamic steady state. To demonstrate this we plot in Fig.~\ref{fig:amplitude_asymmetry} the steady state magnitude of the interface fluctuations versus the curvature asymmetry, for different values of activity and surface tensions. For large fluctuations, the interface is no longer single-valued and fluctuations cannot be quantified in terms of a height field. Instead we measure the local orientation $\langle \theta^2\rangle$ of the interface defined by tangent angle
\begin{equation}
    \theta(s) = \arctan\left(\frac{\delta y(s)}{\delta x(s)}\right)\;.
\end{equation}
 Here the interface is parameterized by $(x(s), y(s))$ and $\delta x = x(s+\Delta s) - x(s), \delta y = y(s +\Delta s) - y(s)$. We then measure the variance $\langle \theta^2 \rangle$ of the tangent angle averaged across the interface and over time. Clearly  $\langle \theta^2 \rangle =0$ for a flat interface.

It is evident from Fig.~\ref{fig:amplitude_asymmetry} that the amplitude of fluctuations correlates strongly with the mean asymmetry of the interface for all values of activity and surface tension. This suggests a universal behavior of the interface that may be understood in terms of a minimal dynamical model of the interface itself. This is, however, beyond the scope of this work. 

\subsection{Role of topological defects}
As briefly discussed in the main text, beyond the effect of the asymmetric vortical flows near the interface, the curvature asymmetry of the interface is also influenced by the distribution of the topological defects in the liquid crystalline phase. The $+1/2$ and $-1/2$ defects tend to asymmetrically deform the interface, and in turn are distributed asymmetrically near the interface \cite{chaithanya2024transport}. We expect the effect of these defects to be more relevant in the case of interfaces with an active ordered nematic phase. 

\begin{figure}[h!]
    \centering    
    \includegraphics[width=\textwidth]{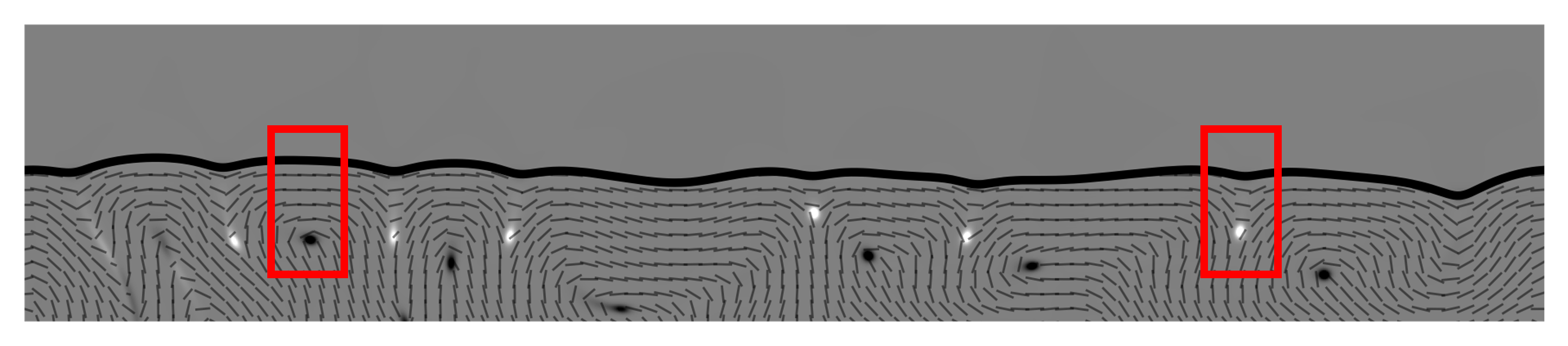}
    \caption{An active (ordered) nematic fluid (bottom) separated from a passive fluid (top); the interface between the two fluids is shown as a black line. The dashed lines show the local nematic director and the grey scale shows the local defect charge density where black (white) corresponds to $+1/2 \; (-1/2)$ defects in the nematic texture \cite{blow2014biphasic}. In red we highlight representative instances where a $-1/2$ defect is closely aligned with the interface where local curvature $\kappa>0$ and a $+1/2$ defect with $\kappa<0$}
    \label{fig:nematicdefects}
\end{figure}

In general, for a bulk active nematic, bend instabilities eventually lead to generation of oppositely charged topological defects. In the presence of a soft interface, we find that the $+1/2$ (comet-shaped, motile) defects tend to nucleate near (and preferentially align with) regions of local negative curvature (the interface peaks) whereas the $-1/2$ (triangular, non-motile) defects tend to coincide with regions of local positive curvature (the interface valleys), as shown in Fig.~\ref{fig:nematicdefects}. Since we expect the effect of these point defects to be less important in the case of the isotropic liquid crystal phase, we do not explore this in detail here. We leave this as an interesting avenue for further study in the context of mixtures of (ordered) active nematic and isotropic passive fluids.

\section{Interface Folding and Invagination of Passive Droplets}
\begin{figure}[h!]
    \centering
    \includegraphics[width=\textwidth]{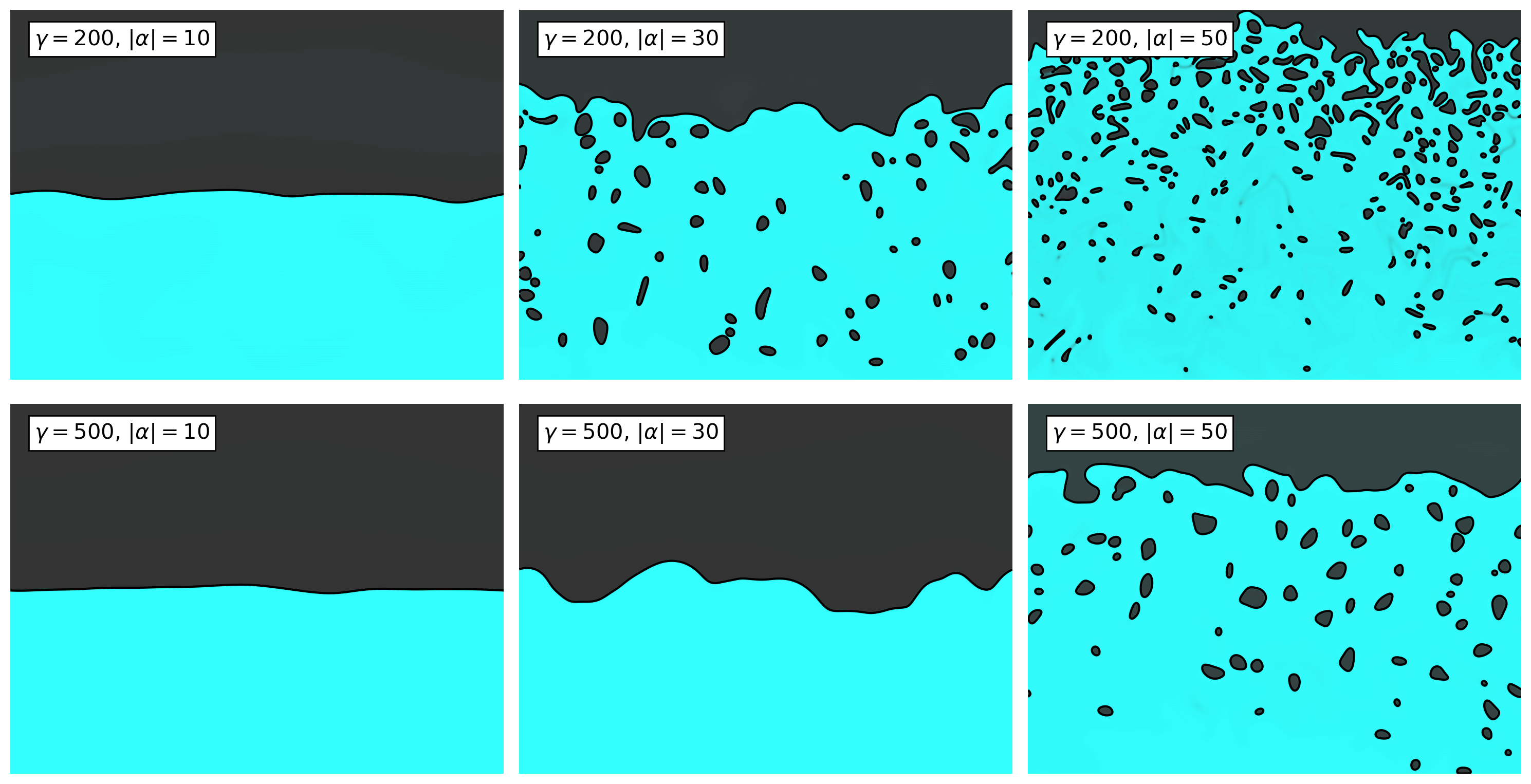}
    \caption{Representative snapshots of the phase field $\phi$ at steady state for increasing activity at two different values of the surface tension.}
    \label{fig:DropletSnapshots}
\end{figure}

\begin{figure}[h!]
    \centering
    \includegraphics[width=\textwidth]{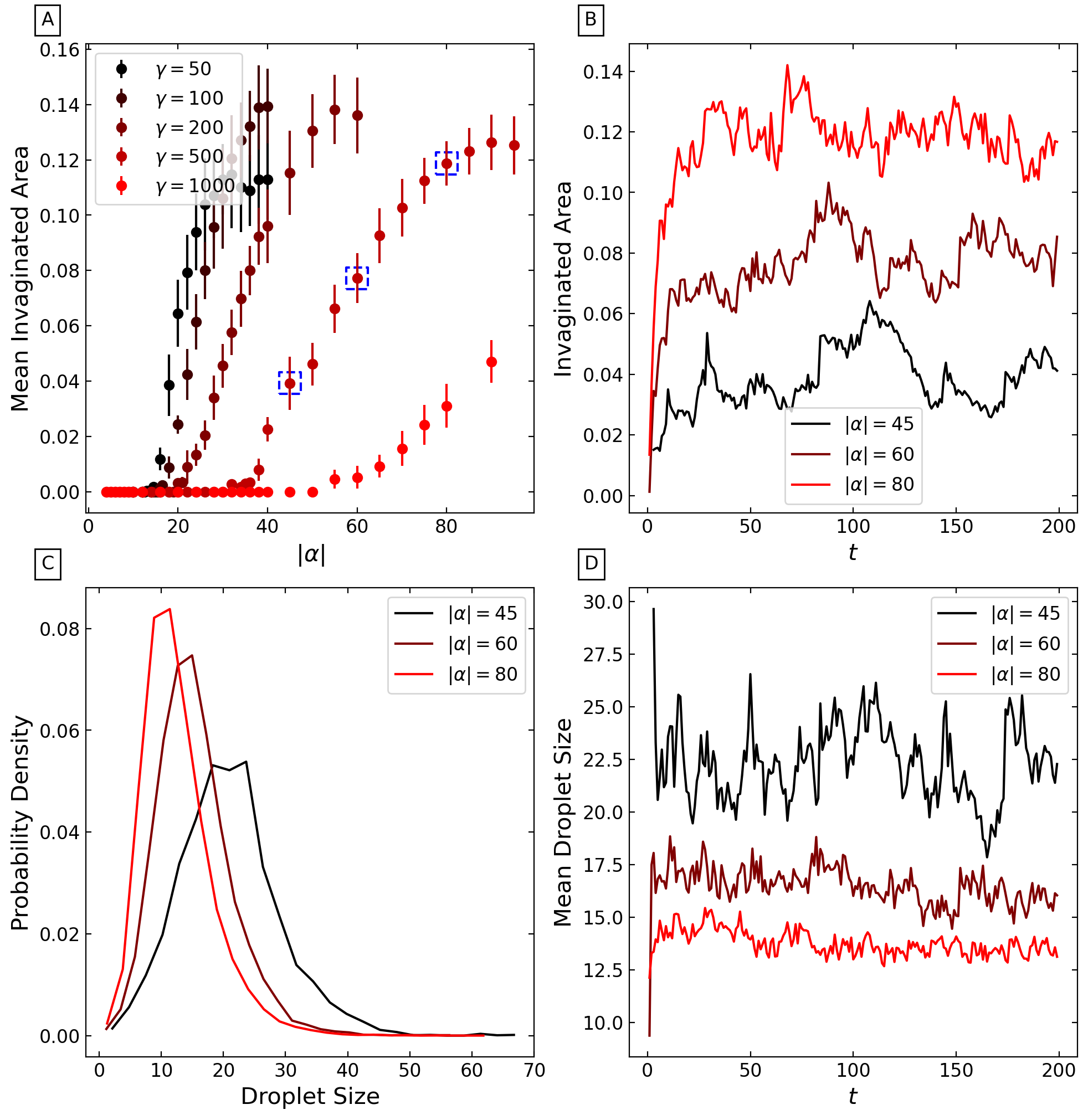}
    \caption{Droplet Formation Statistics: (A) Mean total area (over the steady state) of the passive droplets in the bulk active phase (same as Fig 6). (B) Time series showing the area of the passive droplets over time for the highlighted values of the surface tension and activity (C) Droplet size distributions (over the steady state) for the highlighted points. Here the size of a droplet is defined as $\sqrt{A_d},$ where $A_d$ is the area for each individual droplet (D) Mean size of the droplets, for the highlighted values, as a function of time,}
    \label{fig:Droplet_OP_TimeSeries}
\end{figure}

At large activity, folding of the interface leads to formation of passive droplets invaginated into the bulk active phase (Fig~\ref{fig:DropletSnapshots}). This transition occurs above a critical activity $\alpha_c$ and we use the area of the invaginated droplets as the order parameter to quantify the transition. At any given time, we identify individual droplets 
and quantify their 
size and position as a function of time. From this, we calculate the total invaginated area of the droplets as a function of time. We use this to characterize the steady state and then plot the mean order parameter (Fig.~\ref{fig:Droplet_OP_TimeSeries}A, B). 

We can also quantify the mean droplet size as a function of time, and characterize the distribution of droplet sizes, 
as shown in Fig.~\ref{fig:Droplet_OP_TimeSeries}C, D. 

\clearpage

\begin{figure}[h!]
    \centering
    \vspace{2cm}
    \includegraphics[width=\textwidth]{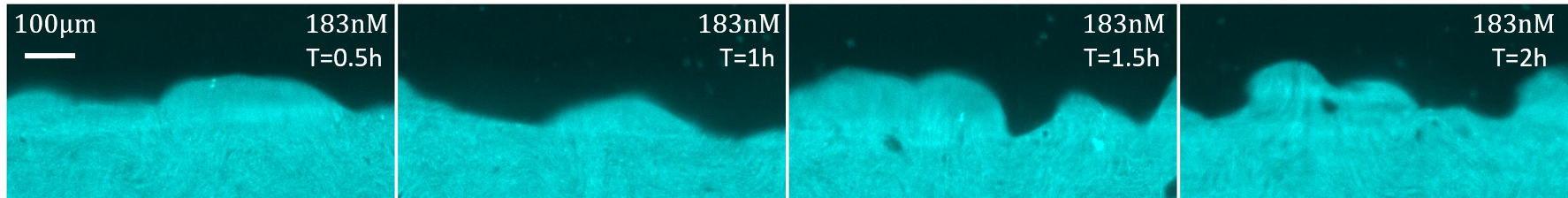}
    \caption{Two-dimensional cross-sections of the interface in a sample at four time points. The Interfacial fluctuation magnitude increased over time. Sample composition: 183nM KSA, 2.1\% PEG and 2.1\% dextran.}
    \label{fig:timeevolution}
\end{figure}

\begin{figure}[h!]
    \centering
    \vspace{2cm}
    \includegraphics[width=\textwidth]{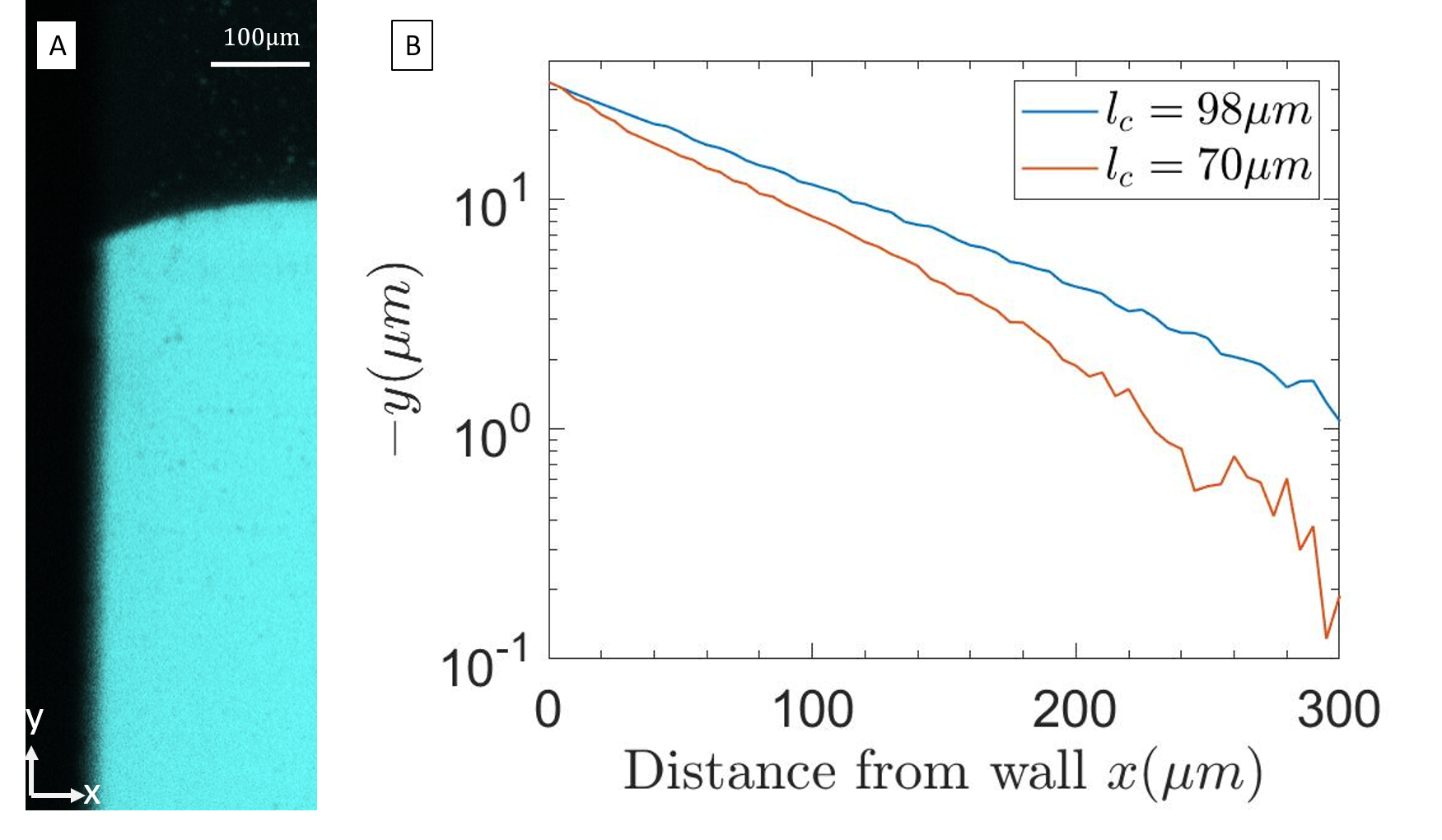}
    \caption{Measurement of capillary lengths between the two separated phases. (A) A two-dimensional crosssection of the interface profile near the wall in a static sample. The dextran phase de-wets the FEP wall. (B) The interface height $y$ versus distance from the wall $x$ plotted on a log-linear scale, demonstrating the exponential dependence of the height near the wall. The capillary lengths of the samples with two different polymer concentrations are 98~$\mu$m and 70~$\mu$m each. The values are obtained by fitting an exponential function to the interface profile near the wall. The measured samples contained no MTs or KSA and contained 2.1\% PEG and 2.1\% dextran ((A), blue curve in (B)) and 2.0\% PEG and 2.0\% dextran (orange curve in (B)). }
    \label{fig:capillary}
\end{figure}

\begin{figure}[h!]
    \centering
    \includegraphics[width=0.7\textwidth]{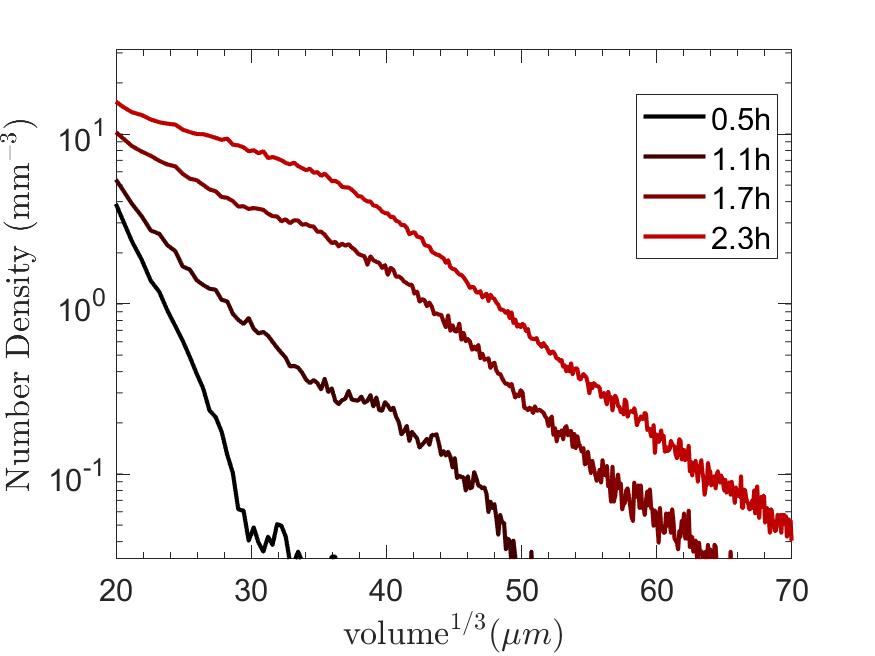}
    \caption{The size of invaginated passive droplets in the active phase grows over time. The figure shows the probability distribution of droplet sizes at four times. Each curve is an average over 0.6h. Sample composition: 183~nM KSA, 2.0\% PEG and 2.0\% dextran.}
    \label{fig:dropletsize}
\end{figure}
\clearpage
\FloatBarrier

\movie{Two-dimensional cross-sections of an active interface imaged over the entire sample lifetime. Interfacial fluctuations increase over time. The inset shows time after the sample was prepared. Sample composition: KSA, 183~nM; 2.1\% PEG and 2.1\% dextran.}

\movie{Two-dimensional cross-sections of the interfaces in four samples with increasing activity (KSA concentration). Interfacial fluctuations increase with increasing KSA. The movie starts at $t=2h$ after the samples were prepared. The inset shows time relative to first frame. Sample composition: 2.1\% PEG, 2.1\% dextran and increasing KSA.}

\movie{Three examples of invagination of passive droplets by the active phase. The valley in the interface grows deeper with time. Eventually, the vertical walls of the deep valley merge together to form a passive droplet that is enveloped by the active fluid. The inset shows time relative to the three starting frames. Sample composition: KSA, 183~nM; 2.1\% PEG and 2.1\% dextran.}
\movie{Three-dimensional view of invaginations happening at the interface. The movie starts at $t=2.7$~h after the sample was prepared. The inset shows time relative to the starting frame. Sample composition: 183~nM KSA, 2.0\% PEG and 2.0\% dextran.}

\movie{Formation of perforated a sample in the high activity and low surface tension regime. Top: two-dimensional cross sections of the active phase being perforated. Bottom: the total volume of enveloped droplets as a function of time. The inset shows time after the sample was prepared. Sample composition: 183~nM KSA, 2.0\% PEG, and 2.0\% dextran.} 

\movie{Passive droplets accumulate at the active-passive interface. The droplets are frequently adjacent to the bulk interface but remain separated by a thin layer of active fluid. Passive droplets are rarely if ever observed to merge with the top bulk phase. Red curve at 4:05 marks an example of such a droplet. The movie starts at $t=2.7$~h after the sample was prepared. The inset shows time relative to the first frame. Sample composition: 183~nM KSA, 2.0\% PEG and 2.0\% dextran.}

\movie{Visualization of passive droplets (red) in the active phase. Both droplet number and size increased over time. The inset shows time after the sample was prepared. Sample composition: 183~nM KSA, 2.0\% PEG and 2.0\% dextran.}

\movie{Simulation videos for the phase field $\phi~(N=1024)$ at a fixed surface tension $\gamma =200.0$, and increasing activities, $\alpha =\{-18.0, -20.0, -24.0\}.$ Here, the videos are only shown for the first half of the total duration i.e until $t=100.$}

\movie{Simulation video for the phase field $\phi~(N=1024)$ at large activity showing the dynamics in the perforated active state, with surface tension $\gamma =200.0,$ and activity $\alpha =-30.0$.}

\end{document}